\documentclass[12pt,a4paper,english]{article}
\usepackage{jheppub}
\usepackage{amsfonts}
\usepackage{mathtools}
\usepackage{braket}
\usepackage{slashed}
\usepackage{multirow}
\usepackage[T1]{fontenc}
\usepackage[latin9]{inputenc}
\usepackage{color}
\usepackage{amsmath}
\usepackage{esint}
\usepackage{amssymb}
\usepackage{graphicx}

\usepackage{pstricks}
\usepackage{color}
\usepackage{bbold}

\newcommand{\be}{\begin{equation}}
\newcommand{\ee}{\end{equation}}
\newcommand{\bea}{\begin{eqnarray}}
\newcommand{\eea}{\end{eqnarray}}

\newcommand{\half}{\tfrac{1}{2}}
\newcommand{\p}{\partial}
\newcommand{\unit}{{\mathbb 1}}

\def\cD{{\cal D}}

\def\cG{{\cal G}}
\def\cL{{\cal L}}
\def\cM{{\cal M}}
\def\cO{{\cal O}}
\def\cP{{\cal P}}
\def\cR{{\cal R}}
\def\cV{{\cal V}}

\newcommand{\gb}{\bar{g}}
\newcommand{\Nb}{\bar{N}}
\newcommand{\Nh}{\hat{N}}
\newcommand{\sib}{\bar{\sigma}}
\newcommand{\sh}{\hat{\sigma}}
\newcommand{\Db}{\bar{D}}
\newcommand{\Kb}{\bar{K}}
\newcommand{\Rb}{\bar{R}}

\newcommand{\Mbb}{{\mathbb{M}}}
\newcommand{\Kbb}{{\mathbb{K}}}
\newcommand{\Pbb}{{\mathbb{P}}}
\newcommand{\Vbb}{{\mathbb{V}}}

\usepackage{babel}

\title{Renormalization group fixed points of foliated gravity-matter systems}

\author{Jorn Biemans$^1$,}
\author{Alessia Platania$^{1,2}$,}
\author{Frank Saueressig$^1$}
\affiliation{
$^1$ Institute for Mathematics, Astrophysics and Particle Physics (IMAPP),\\
Radboud University Nijmegen, Heyendaalseweg 135, 6525 AJ Nijmegen, The Netherlands\\ 
}
\affiliation{
$^2$ Department of Physics and Astronomy, University of Catania, Via S. Sofia 63, 95123 Catania, Italy\\ INFN, Catania Section, via S. Sofia 64, 95123, Catania, Italy \\ INAF, Catania Astrophysical Observatory, via S. Sofia 78, 95123, Catania, Italy \\
}
\emailAdd{jbiemans@science.ru.nl}
\emailAdd{alessia.platania@oact.inaf.it}
\emailAdd{f.saueressig@science.ru.nl}

\abstract{
We employ the Arnowitt-Deser-Misner formalism to study the renormalization group flow of gravity minimally coupled to an arbitrary number of scalar, vector, and Dirac fields. The decomposition of the gravitational degrees of freedom into a lapse function, shift vector, and spatial metric equips spacetime with a preferred (Euclidean) ``time''-direction. In this work, we provide a detailed derivation of the renormalization group flow of Newton's constant and the cosmological constant on a flat Friedmann-Robertson-Walker background. Adding matter fields, it is shown that their contribution to the flow is the same as in the covariant formulation and can be captured by two parameters $d_g$, $d_\lambda$. We classify the resulting fixed point structure as a function of these parameters finding that the existence of non-Gaussian renormalization group fixed points is rather generic. In particular the matter content of the standard model and its most common extensions gives rise to one non-Gaussian fixed point with real critical exponents suitable for Asymptotic Safety. Moreover, we find non-Gaussian fixed points for any number of scalar matter fields, making the scenario attractive for cosmological model building.
}


\begin{document}
\maketitle


\section{Introduction}
\label{intro}
The quantization of the gravitational force is an outstanding open problem in theoretical high-energy physics. 
In this context, Asymptotic Safety, first proposed by Weinberg \cite{Weinberg:1980gg,Weinproc1} and recently reviewed in \cite{Niedermaier:2006wt,Codello:2008vh,Litim:2011cp,Percacci:2011fr,Reuter:2012id,Reuter:2012xf,Nagy:2012ef}, may provide an attractive mechanism for obtaining a consistent and predictive quantum theory for gravity within the framework of quantum field theory. Given that asymptotic safety is a rather general concept whose applicability is not limited to the gravitational interactions it was soon realized that this mechanism may also be operative once the gravitational degrees of freedom are supplemented by matter fields \cite{Percacci:2002ie,Percacci:2003jz}. In this way the asymptotic safety scenario may also provide a framework for unifying all fundamental forces and matter fields populating the universe within a single quantum field theory.

The key ingredient underlying the asymptotic safety mechanism is a renormalization group (RG) fixed point which controls the behavior of the theory at ultra-high energies. The fixed point then ensures that all dimensionless coupling constants remain finite, preventing the occurrence of unphysical UV divergences. Provided that it also comes with a finite number of eigendirections along which the flow is dragged into the fixed point for increasing energy this construction has the same predictive power as a perturbatively renormalizable quantum field theory. For the case where 
the gravitational degrees of freedom are encoded in fluctuations of the (Euclidean) spacetime metric, defining the so-called metric approach to Asymptotic Safety, the
existence of a suitable non-Gaussian fixed point (NGFP) has been demonstrated in a vast number of works including the projection of the gravitational RG flow onto the Einstein-Hilbert action \cite{Reuter:1996cp,Lauscher:2001ya,Reuter:2001ag,Litim:2003vp,Fischer:2006fz,Donkin:2012ud,Nagy:2013hka},  
$f(R)$-type actions build from finite polynomials constructed from the curvature scalar $R$ \cite{Lauscher:2002sq,Codello:2007bd,Machado:2007ea,Codello:2008vh,Falls:2013bv,Falls:2014tra,Demmel:2014hla,Falls:2016msz}, and including the square of the Weyl tensor \cite{Codello:2006in,Benedetti:2009rx,Benedetti:2009gn,Groh:2011vn}. Moreover, Ref.\ \cite{Gies:2016con}  established that this NGFP also persists once the perturbative two-loop counterterm found by Goroff and Sagnotti \cite{Goroff:1985th} is included in the projection. A complementary class of approximations which also keeps track of the fluctuation fields, corroborates this picture \cite{Manrique:2009uh,Manrique:2010mq,Manrique:2010am,Christiansen:2012rx,Codello:2013fpa,Christiansen:2014raa,Becker:2014qya,Christiansen:2015rva}. Moreover, approximations including an infinite number of scale-dependent coupling constants are currently under development \cite{Codello:2010mj,Benedetti:2012dx,Demmel:2012ub,Dietz:2012ic,Demmel:2013myx,Dietz:2013sba,Benedetti:2013nya,Demmel:2014sga,Percacci:2015wwa,Borchardt:2015rxa,Demmel:2015oqa,Ohta:2015efa,Ohta:2015fcu,Henz:2016aoh,Labus:2016lkh,Dietz:2016gzg}. Starting from \cite{Percacci:2002ie,Percacci:2003jz} it has also been shown that the  Asymptotic Safety mechanism may also play a key role in the high-energy completion of a large class of gravity-matter models \cite{Zanusso:2009bs,Vacca:2010mj,Harst:2011zx,Eichhorn:2011pc,Dona:2013qba,Dona:2014pla,Meibohm:2015twa,Oda:2015sma,Dona:2015tnf,Meibohm:2016mkp,Eichhorn:2016esv,Henz:2016aoh}. 

An open question in the metric approach to Asymptotic Safety is the so-called ``problem of time'' (see \cite{Isham:1992ms} for review). While quantum mechanics and quantum field theory in a fixed Minkowski background possess a natural notion of time, the notion of time in a dynamical (and possibly fluctuating) spacetime becomes rather involved. A way to address this question in general relativity is the Arnowitt-Deser-Misner (ADM)-formalism. In this case the spacetime metric is decomposed into a Lapse function $N$, a shift vector $N_i$, and a metric $\sigma_{ij}$ which measures distances on the spatial slices $\Sigma_t$ defined as hypersurfaces where the time-variable $t$ is constant. The foliation structure then leads to natural time-direction.

A  functional renormalization group equation (FRGE) for the effective average action \cite{Wetterich:1992yh,Morris:1993qb,Reuter:1993kw,Reuter:1996cp} tailored to the ADM-formalism has been constructed in \cite{Manrique:2011jc,Rechenberger:2012dt}. A first evaluation of the resulting RG flow within the Matsubara-formalism provided strong indications that the UV fixed point underlying the Asymptotic Safety program is robust under a change from Euclidean to Lorentzian signature \cite{Manrique:2011jc,Rechenberger:2012dt}. Moreover, the FRGE in ADM-variables provides a powerful tool for studying RG flows within Ho\v{r}ava-Lifshitz gravity \cite{Horava:2009uw} since it allows for anisotropic scaling relations between spatial and time-directions by including higher-derivative intrinsic curvature terms \cite{Contillo:2013fua,D'Odorico:2014iha,D'Odorico:2015yaa}.
 
The purpose of the present work is twofold. Firstly, it provides all technical details underlying the construction of RG flows from the ADM-formalism on a flat Friedmann-Robertson-Walker background studied in \cite{Biemans:2016rvp}. Here the key ingredient is the gauge-fixing scheme presented in Sect.\ \ref{sect.3a} which leads to regular propagators for \emph{all} component fields including the lapse function and the shift vector thereby avoiding the pathologies encountered in temporal gauge. Secondly, we initiate the study of matter effects in this setting, by computing the scale-dependence of Newton's constant $G_k$ and the cosmological constant $\Lambda_k$ for foliated gravity-matter systems containing an arbitrary number of minimally coupled scalars, $N_S$, vector fields $N_V$, and Dirac fermions $N_D$. The inclusion of the matter fields leads to a two-parameter deformation of the beta functions controlling the flow of $G_k$ and $\Lambda_k$ in the pure gravity case. Analyzing the beta functions of the gravity-matter systems utilizing these deformation parameters allows classifying their fixed point structure of the model independently of a specific choice of regulator in the matter sector. The fixed point structure found for a specific gravity-matter model can then determined by evaluating the map relating its field content to the deformation parameters. In particular, we find that the matter content of the standard model of particle physics as well as many of its phenomenologically motivated extensions are located in areas which give rise to a single UV fixed point with real critical exponents.  These findings provide a first indication that the asymptotic safety mechanism encountered in the case of pure gravity may carry over to the case of gravity-matter models with a realistic matter field content, also in the case where spacetime is equipped with a foliation structure.

This work is organized as follows. Sect.\ \ref{sect.2} reviews the ADM-formalism and the construction of the corresponding FRGE \cite{Manrique:2011jc,Rechenberger:2012dt}. Our ansatz for the effective average action and the evaluation of the resulting RG flow on a (Euclidean) Friedmann-Robertson-Walker (FRW) background is presented in Sect.\ \ref{sect.3}. In particular, Sect.\ \ref{sect.3a} summarizes the construction of our novel gauge-fixing scheme leading to regular propagators for all component fields. Limiting the analysis to $D=3+1$ spacetime dimensions, the  flow equation in the context of pure gravity are analyzed in Sect.\ \ref{sect.42} while the fixed point structure appearing in gravity-matter systems is discussed in Sect.\ \ref{sect.43}.
 We provide a short summary and discussion of our findings in Sect.\ \ref{sect.7}. Technical details about the background geometry, the construction of the Hessians in a flat Friedmann-Robertson-Walker background, and the evaluation of the operator traces entering the FRGE are relegated to the App.\ \ref{App.A}, App.\ \ref{App.B}, and App.\ \ref{App.C}, respectively.

\section{Renormalization group flows on foliated spacetimes}
\label{sect.2}
The functional renormalization group equation on foliated spacetimes has been constructed in \cite{Manrique:2011jc,Rechenberger:2012dt} and we review the formalism in the following section. For a pedagogical introduction to the $3+1$-formalism the reader is referred to \cite{Gourgoulhon:2007ue}.
\subsection{Arnowitt-Deser-Misner decomposition of spacetime}
\label{sect.2a}
We start from a $D$-dimensional Euclidean manifold $\cM$ with metric $\gamma_{\mu\nu}$, carrying coordinates $x^\alpha$. In order to be able to perform a Wick rotation to Lorentzian signature, we define a time function $\tau(x)$ which assigns a specific time $\tau$ to each spacetime point $x$. This can be used to decompose $\cM$ into a stack of spatial slices $\Sigma_{\tau_i} \equiv \left\{ x: \tau(x) = \tau_i \right\}$ encompassing all points $x$ with the same value of the ``time-coordinate'' $\tau_i$. The gradient of the time function $\p_\mu \tau$ can be used to define a vector $n^\mu$ normal to the spatial slices, $n_\mu \equiv N \p_\mu \tau$ where the lapse function $N(\tau,y^i)$ is used to ensures the normalization $\gamma_{\mu\nu} \, n^\mu \, n^\nu = 1$. Furthermore, the gradient can be used to introduce a vector field $t^\mu$ satisfying $t^\mu \p_\mu \tau = 1$. Denoting the coordinates on $\Sigma_\tau$ by $y^i$, $i = 1, \ldots, d$ the tangent space on a point in $\cM$ can then be decomposed into the space tangent to $\Sigma_\tau$ and its complement. The corresponding basis vectors can be constructed from the Jacobians
\be\label{proj1}
t^\mu = \left . \frac{\partial x^\mu}{\partial \tau} \right|_{y^i} \, , \qquad e_i{}^\mu = \left. \frac{\p x^\mu}{\p y^i} \right|_\tau \, . 
\ee
The normal vector then satisfies $\gamma_{\mu\nu} \, n^\mu \, e_i{}^\nu = 0$.

The spatial coordinate systems on neighboring spatial slices can be connected by constructing the integral curves $\gamma$ of $t^\mu$ and requiring that $y^i$ is constant along these curves. A priori $t^\mu$ is neither tangent nor orthogonal to the spatial slices. Using the Jacobians \eqref{proj1} it can be decomposed into its components normal and tangent to $\Sigma$
\be\label{tdec}
t^\mu = N \, n^\mu + N^i \, e_i{}^\mu \, .
\ee
where $N^i(\tau,y^i)$ is called the shift vector. Analogously, the coordinate one-forms transform according to
\be
dx^\mu = t^\mu d\tau + e_i{}^\mu dy^i = N n^\mu d\tau + e_i{}^\mu \, (dy^i + N^i d\tau) \, . 
\ee
Defining the metric on the spatial slice $\sigma_{ij} = e_i{}^\mu \, e_j{}^\nu \, \gamma_{\mu\nu}$ the line-element $ds^2 = \gamma_{\mu\nu} \, dx^\mu dx^\nu$ written in terms of the ADM fields
takes the form 
\be\label{fol1}
ds^2 = \gamma_{\alpha\beta} \, dx^\alpha dx^\beta 
=  N^2 d\tau^2 +  \sigma_{ij} \, (dy^i + N^i d \tau) 
(dy^j +  N^j d \tau) \, . 
\ee 
Note that in this case the lapse function $N$, the shift vector $N^i$ and the induced metric on the spatial slices $\sigma_{ij}$ depend on the spacetime coordinates $(\tau, y^i)$.\footnote{This situation differs from projectable Ho\v{r}ava-Lifshitz gravity where $N$ is restricted to be a function of (Euclidean) time $\tau$ only.} In terms of metric components, the decomposition \eqref{fol1} implies
\be\label{metcomp}
\gamma_{\alpha\beta} = \left(
\begin{array}{cc}
	N^2 + N_i N^i \; \;  & \;  \; N_j \\
	N_i &  \sigma_{ij} 
	\end{array}
\right) \, , \qquad 
\gamma^{\alpha\beta} = \left(
\begin{array}{cc}
 \frac{1}{N^{2}} \; \;  & \;  \; -  \frac{N^j}{ N^{2}}  \\
 -  \frac{N^i}{ N^{2}}  	 \; \;  & \;  \;  \sigma^{ij} +  \, \frac{ N^i \,  N^j}{ N^{2}} 
\end{array}
\right) \,
\ee
where spatial indices $i,j$ are raised and lowered with the metric on the spatial slices. 

An infinitesimal coordinate transformation $v^\alpha(\tau,y)$ acting on the metric can be expressed in terms of the Lie derivative $\cL_v$
\be\label{diffeo1}
\delta \gamma_{\alpha\beta} = \cL_v \, \gamma_{\alpha\beta} \, . 
\ee 
Decomposing 
\be\label{vdec}
v^\alpha = \left(f(\tau,y), \zeta^i(\tau,y)\right)
\ee
 into its temporal and spatial parts, the transformation \eqref{diffeo1} determines the transformation properties of the component fields under Diff($\cM$)
\be\label{eq:gaugeVariations}
\begin{split}
\delta N &= \p_\tau (f N ) + \zeta^k \p_k N  - N N^i\p_i f \, , \\
\delta  N_i &= \partial_\tau( N_i f) + \zeta^k\partial_k  N_i +  N_k\partial_i\zeta^k
+ \sigma_{ki}\partial_\tau \zeta^k
+  N_k  N^k\partial_i f  +  N^2\partial_i f \, , \\
\delta\sigma_{ij} &= f\p_\tau \sigma_{ij} + \zeta^k\p_k \sigma_{ij} + \sigma_{jk}\p_i\zeta^k + \sigma_{ik}\p_j\zeta^k + N_j\p_i f + N_i\p_j f  \, . 
\end{split}
\ee
For completeness, we note 
\be\label{Nui}
\delta N^i = \p_\tau(N^i f) + \zeta^j\p_j N^i - N^j\p_j \zeta^i + \p_\tau \zeta^i - N^i N^j \p_j f  + N^2 \sigma^{ij}\p_j f \, .
\ee
Denoting expressions in Euclidean and Lorentzian signature by subscripts $E$ and $L$, the Wick rotation is implemented by
\be\label{Wickrot}
\tau_E \rightarrow - i \tau_L \, , \qquad N^i_E \rightarrow i N^i_L \, . 
\ee

The (Euclidean) Einstein-Hilbert action written in ADM fields reads
\be\label{EHaction}
S^{\rm EH} = \frac{1}{16 \pi G} \int d\tau d^dy \, N \sqrt{\sigma} \left[ K_{ij} \, \cG^{ij,kl} \, K_{kl} - {}^{(d)}R + 2 \Lambda \right] \, .
\ee
Here ${}^{(d)}R$ denotes the intrinsic curvature on the $d$-dimensional spatial slice,
\be\label{Kext}
K_{ij} \equiv \frac{1}{2 N} \left( \p_\tau \sigma_{ij} - D_i N_j - D_j N_i \right) \, , \quad K \equiv \sigma^{ij} K_{ij}
\ee
are the extrinsic curvature and its trace, and $D_i$ denotes the covariant derivative constructed from $\sigma_{ij}$. The kinetic term is determined by the Wheeler-de Witt metric
\be
\cG^{ij,kl} \equiv \sigma^{ik} \, \sigma^{jl} - \lambda \, \sigma^{ij} \, \sigma^{kl} \, . 
\ee
The parameter $\lambda = 1$ is fixed by requiring invariance of the action with respect to Diff($\cM$) and we adhere to this value for the rest of this work. 

When studying the effects of matter fields in Sect.\ \ref{sect.43}, we supplement the gravitational action \eqref{EHaction} by $N_S$ scalar fields, $N_V$ abelian gauge fields and $N_D$ Dirac fields minimally coupled to gravity
\be
S^{\rm matter} = S^{\rm scalar} + S^{\rm vector} + S^{\rm fermion} \, , 
\ee 
where
\be\label{matter}
\begin{split}
S^{\rm scalar} = & \frac{1}{2} \sum_{i=1}^{N_S} \int d\tau d^dx N \sqrt{\sigma} \left[ \, \phi^i \, \Delta_0 \, \phi^i \right] \, , \\
S^{\rm vector} = & \frac{1}{4} \sum_{i=1}^{N_V} \int d\tau d^dx N \sqrt{\sigma} \left[ \, g^{\mu\nu} g^{\alpha\beta} F_{\mu\alpha}^i F_{\nu\beta}^i \right] \,  + \frac{1}{2\xi} \sum_{i=1}^{N_V} \int d\tau d^dx \Nb \sqrt{\sib} \left[ \, \gb^{\mu\nu} \Db_\mu A_\nu^i \right]^2 \\  
& + \sum_{i=1}^{N_V} \int d\tau d^dx \Nb \sqrt{\sib} \left[ \, \bar{C}^i \, \Delta_0 \, C^i \, \right] \, , \\
 S^{\rm fermion} = & \, i \, \sum_{i=1}^{N_D} \int d\tau d^dx N \sqrt{\sigma} \left[ \bar{\psi}^i \, \slashed{\nabla} \, \psi^i \right] \, .   
\end{split}
\ee
The summation index $i$ runs over the matter species and we adopt Feynman gauge setting $\xi=1$. In the context of Asymptotic Safety, matter sectors of this type have been discussed in the context of the covariant approach in \cite{Percacci:2002ie,Percacci:2003jz} with extensions considered recently in \cite{Dona:2013qba,Labus:2015ska,Dona:2015tnf,Meibohm:2015twa}. In particular, our treatment of the Dirac fermions follows \cite{Dona:2012am,Dona:2013qba}. 
All matter actions are readily converted to by using the projector \eqref{proj1}. In order to retain compact expressions, we refrain from giving this decomposition explicitly, though.  
 
\subsection{Functional renormalization group equation}
\label{sect.2b}
The first step in deriving the FRGE for the effective average action $\Gamma_k$ \cite{Wetterich:1992yh,Morris:1993qb,Reuter:1993kw,Reuter:1996cp}
specifies the field content of the model. For foliated spacetimes, it is natural to encode the gravitational degrees of freedom in terms of the ADM-fields $\{N, N_i, \sigma_{ij}\}$. Additional matter degrees of freedom
are easily incorporated by including additional fields in the construction. 
The construction of $\Gamma_k$ makes manifest use of the background field method. Following \cite{Rechenberger:2012dt} we use a linear split of the ADM fields into background fields (marked with an bar) and fluctuations (indicated by a hat)\footnote{Strictly speaking, the fields appearing in the effective average action are the vacuum expectation values of the classical fields introduced in the previous subsection. In order to keep our notation light, we use the same notation for both fields, expecting that the precise meaning is clear from the context.}
\be\label{linearsplit}
N = \Nb + \Nh \, , \qquad N_i = \Nb_i + \Nh_i \, , \qquad \sigma_{ij} = \sib_{ij} + \sh_{ij} \, .
\ee 
Conveniently, we will denote the sets of physical fields, background fields and fluctuations by $\chi$, $\bar{\chi}$, and $\hat{\chi}$, respectively. I.e., $\chi = \{N, N_i, \sigma_{ij}, \ldots \}$ where the dots indicate ghost fields and potentially additional matter fields.

The effective average action is then obtained in the usual way. Starting from a generic diffeomorphism invariant action $S^{\rm grav}[N,N_i,\sigma_{ij}]$, one formally writes down the generating functional
\be
Z_k[J; \bar{\chi}] \equiv \int \cD \Nh \cD \Nh_i \cD \hat{\sigma} \, \exp\left[-S^{\rm grav} - S^{\rm gf} - S^{\rm ghost} - \Delta_kS - S^{\rm source}\right] \, , 
\ee
where $S^{\rm grav}$ is supplemented by a suitable gauge-fixing term $S^{\rm gf}$, a corresponding ghost action $S^{\rm ghost}$ exponentiating the Faddeev-Popov determinant, and source terms $S^{\rm source}$ for the fluctuation fields. The crucial ingredient is the infrared regulator 
\be\label{regulator}
\Delta_k S \equiv \half \int d\tau d^dy \sqrt{\sib} \Nb \, \left[ \, \hat{\chi} \, \cR_k[\bar{\chi}] \, \hat{\chi} \, \right] \, ,  
\ee
where the matrix-valued kernel $\cR_k[\bar{\chi}]$ is constructed from the background metric and provides a scale-dependent mass term for fluctuations with momenta $p^2 \lesssim k^2$. Based on the partition function, we define the generating functional for the connected Green functions
\be
W_k[J; \bar{\chi}] \equiv \log\left[ Z_k \right] \, . 
\ee
The effective average action is then obtained as
\be
\Gamma_k[\hat{\chi};\bar{\chi}] \equiv \widetilde \Gamma_k[\hat{\chi};\bar{\chi}] - \Delta_kS[\chi;\bar{\chi}] 
\ee
where $\widetilde \Gamma_k$ is the Legendre-transform of $W_k$. In general $\Gamma_k$ consists of a (generic) gravitational action $\Gamma_k^{\rm grav}$ supplemented by a suitable gauge-fixing $\Gamma_k^{\rm gf}$, ghost action $\Gamma_k^{\rm ghost}$ and, potentially, a matter action 
\be\label{Gform}
\Gamma_k = \Gamma_k^{\rm grav} + \Gamma_k^{\rm gf} + \Gamma_k^{\rm ghost} + \Gamma_k^{\rm matter} \, , 
\ee
with the gauge-fixing constructed from the background field method.

The key property of $\Gamma_k$ is that its scale-dependence is governed by a formally exact FRGE 
\be\label{FRGE}
k \p_k \Gamma_k = \frac{1}{2} \, {\rm STr} \left[ \left( \Gamma_k^{(2)} + \cR_k \right)^{-1} \, k \p_k \cR_k \right] \, . 
\ee
Here $\Gamma_k^{(2)}$ denotes the second variation of $\Gamma_k$ with respect to the fluctuation fields $\hat{\chi}$, STr contains a graded sum over component fields and an integration over loop momenta, and the matrix-valued IR regulator $\cR_k$ has been introduced in \eqref{regulator}. The interplay between the regularized propagator $\left( \Gamma_k^{(2)} + \cR_k \right)^{-1}$ and $k \p_k \cR_k$ ensures that the right-hand-side of the FRGE is actually finite. Moreover, the FRGE realizes Wilson's idea of renormalization in the sense that the  flow of $\Gamma_k$ is essentially driven by fluctuations located in a small momentum-interval situated at the RG scale $k$. 

At this stage, the following remark is in order. Owed to the non-linearity of the ADM decomposition, the transformation of the ADM fields under the full diffeomorphism group is non-linear. In combination with the linear split \eqref{linearsplit} this entails that $S^{\rm gf}$ and $\Delta_kS$ preserve a subgroup of the full diffeomorphism group as a background symmetry only.  Inspecting eqs.\ \eqref{eq:gaugeVariations} and \eqref{Nui} one sees, that restricting $f = f(\tau)$ eliminates the quadratic terms in the transformations laws. This indicates that the flow equation \eqref{FRGE} respects foliation preserving diffeomorphisms where, by definition, the transformation $f = f(\tau)$ is independent of the spatial coordinates. Also see \cite{Rechenberger:2012dt} for a detailed discussion.

\section{RG flows on a Friedmann-Robertson-Walker background}
\label{sect.3}
In this section we use the FRGE \eqref{FRGE} to determine the beta functions encoding the scale-dependence of Newton's constant and the cosmological constant in the context of pure gravity and gravity minimally coupled to non-interacting matter fields. The key ingredient in the construction is a novel gauge-fixing scheme introduced in Sect.\ \ref{sect.3a} where all ADM-fields acquire a relativistic dispersion relation. Our discussion primarily focuses on the gravitational sector of the flow, incorporating the contributions from the matter sector at the very end only.
%
\subsection{The Einstein-Hilbert ansatz}
\label{sect.3mod}
Finding exact solutions of the FRGE \eqref{FRGE} is rather difficult. A standard way of constructing approximate solutions, which does not rely on the expansion in a small coupling constant, is to restrict the interaction monomials in $\Gamma_k$ to a specific subset and subsequently project the RG flow onto the subspace spanned by the ansatz. 
In the present work, we will project the full RG flow onto the  Einstein-Hilbert action written in terms of the ADM-fields
\be\label{GammaEH}
\Gamma_k^{\rm grav} \simeq \frac{1}{16 \pi G_k} \int d\tau d^dy \, N \sqrt{\sigma} \left[ K_{ij} K^{ij} - K^2 - {}^{(d)}R + 2 \Lambda_k \right] \, .
\ee
This ansatz contains two scale-dependent couplings, Newton's constant $G_k$ and the cosmological constant $\Lambda_k$. Their scale-dependence can be read off from the coefficient multiplying the square of the extrinsic curvature and the spacetime volume, respectively. 

In order to facilitate the computation, it then suffices to work out the flow on a background which allows to distinguish between these two interaction monomials. For the ansatz \eqref{GammaEH} it then suffices to evaluate the flow on a flat (Euclidean)  Friedmann-Robertson-Walker (FRW) background
\be\label{FRWback}
\gb_{\mu\nu} = {\rm diag} \left[ \, 1 \, , \, a(\tau)^2 \, \delta_{ij}\right] \qquad \Longleftrightarrow \qquad 
\Nb = 1 \, , \quad \Nb_i = 0 \, , \quad  \sib_{ij} = a(\tau)^2 \,  \delta_{ij} \, ,  
\ee
where $a(\tau)$ is a positive, time-dependent scale factor. Evaluating \eqref{GammaEH} on this background using \eqref{curvatures} yields
\be\label{FRGElhs}
\left. \Gamma_k^{\rm grav} \right|_{\hat{\chi} = 0} 
= \frac{1}{16 \pi G_k} \int d\tau d^dy \, \sqrt{\sib} \left[ - \tfrac{d-1}{d} \, \Kb^2 + 2 \Lambda_k \right] \, ,
\ee
where $\hat{\chi}$ denotes the set of all fluctuation fields. Thus the choice \eqref{FRWback} is sufficiently general to distinguish the two interaction monomials encoding the flow of $G_k$ and $\Lambda_k$. Note that we have not assumed that the background is compact. In particular the ``time-coordinate'' $\tau$ may be taken as non-compact.

\subsection{Hessians, gauge-fixing, and ghost action}
\label{sect.3a}
Constructing the right-hand-side of the flow equation requires the Hessian $\Gamma_k^{(2)}$. Starting with the contribution originating from $\Gamma_k^{\rm grav}$ it is convenient to introduce the building
blocks
\be\label{imon}
\renewcommand{\arraystretch}{1.2}
\begin{array}{ll}
	I_1 \equiv \int d\tau d^dy \, N \sqrt{\sigma} \, K_{ij} K^{ij} \, , \qquad &
	I_2 \equiv \int d\tau d^dy \, N \sqrt{\sigma} \, K^2 \, , \\
	I_3 \equiv \int d\tau d^dy \, N \sqrt{\sigma} \; {}^{(d)}R \, , \qquad &
	I_4 \equiv \int d\tau d^dy \, N \sqrt{\sigma} \, , \\
\end{array}
\ee
such that
\be
\Gamma_k^{\rm grav} = \frac{1}{16 \pi G_k} \left( I_1 - I_2 - I_3 + 2 \Lambda_k \, I_4 \right) \, . 
\ee
Expanding this expression around the background \eqref{FRWback},  
the terms quadratic in the fluctuation fields then take the form
\be\label{GammaEH2}
\delta^2 \Gamma_k^{\rm grav} = \frac{1}{16 \pi G_k} \left( \delta^2 I_1 - \delta^2 I_2 - \delta^2 I_3 + 2 \Lambda_k \, \delta^2 I_4 \right) \, , 
\ee
with the explicit expressions for $\delta^2 I_i$ given in \eqref{varI1}. 

The FRW-background then makes it convenient to express the fluctuation fields in terms of the component fields used in 
cosmic perturbation theory (see, e.g., \cite{Baumann:2009ds} for a pedagogical introduction).
Defining $\Delta \equiv - \sib^{ij} \p_i \p_j$, the shift vector is decomposed into its transverse and longitudinal parts according to
\be\label{TTshift}
\Nh_i = u_i + \p_i \, \tfrac{1}{\sqrt{\Delta}} \, B \, , \qquad \p^i \, u_i = 0 \, . 
\ee 
The metric fluctuations are written as 
\be\label{TTmet}
\sh_{ij} = h_{ij} - \left( \sib_{ij} + \p_i \p_j \, \tfrac{1}{\Delta} \right)  \psi + \p_i \p_j \, \tfrac{1}{\Delta} \, E + \p_i \tfrac{1}{\sqrt{\Delta}} v_j + \p_j \, \tfrac{1}{\sqrt{\Delta}} \, v_i \, , \quad \sh \equiv \sib^{ij} \sh_{ij} \, ,  
\ee
with the component fields subject to the  differential constraints
\be
\p^i \, h_{ij} = 0 \, , \qquad \sib^{ij} h_{ij} = 0 \, , \qquad \p^i v_i = 0 \, . 
\ee
The result obtained from substituting these decompositions into eq.\ \eqref{varI1} is given in eqs.\ \eqref{I1res}, \eqref{I2res}, and \eqref{I3res}. On this basis it is then rather straightforward to write down the explicit form of \eqref{GammaEH2} in terms of the component fields.

At this stage it is instructive to investigate the matrix elements of $\delta^2 \Gamma^{\rm grav}_k$ on flat Euclidean space, obtained by setting $\Kb = 0$. 
 The result is summarized in the second column of Table \ref{Tab.1}.
\begin{table}[t]
	\renewcommand{\arraystretch}{1.4}
	\begin{center}
	\begin{tabular}{|c|l|l|} \hline \hline
		Index 	& matrix element  $32 \pi G_k \, \delta^2\Gamma_k^{{\rm grav}}$ & matrix element $32 \pi G_k \, \left(\delta^2\Gamma_k^{{\rm grav}} + \Gamma_k^{\rm gf} \right)$ \\ \hline
		$h \, h$ & $\Box - 2 \Lambda_k$ & $\Box - 2 \Lambda_k$ \\ \hline
		$v \, v$ & $ 2 \big[-\p_\tau^2 - 2 \Lambda_k \big]$ & $\Box - 2 \Lambda_k$ \\ \hline
		$E \, E$ & $-  \Lambda_k$ & $\half(\Box - 2 \Lambda_k)$ \\ \hline
		$\psi \, \psi$ & $ -  (d-1) (d-2) \big[  \Box - \tfrac{d-3}{d-2} \, \Lambda_k \big] $ & $ - \tfrac{(d-1) (d-3)}{2} \big[  \Box - 2 \, \Lambda_k \big] $ \\ \hline
		$\psi \, E$ & $ - (d-1) \big[-\p_\tau^2 - 2 \Lambda_k \big]$
		& $ - (d-1) \big[\Box - 2 \Lambda_k \big]$ \\ \hline \hline
		$ u \, u$ & $  2 \, \Delta$ & $  2 \, \Box $  \\ \hline
		$ u \, v $ & $-2 \, \p_\tau \sqrt{\Delta}$ & 0  \\ \hline
		$ B \, \psi $ & $2 \, (d-1)  \sqrt{\Delta} \, \p_\tau$ & 0 \\ \hline
		$ \Nh \, \psi $ & $2 \, (d-1)  \big[  \Delta - \Lambda_k \big]$ & $ \, (d-1)  \big[  \Box - 2 \Lambda_k \big]$\\ \hline
		$ \Nh \, E $ & $ - 2 \, \Lambda_k$ & $ \Box - 2 \Lambda_k$ 
		\\ \hline
		$\Nh \Nh$ & $0$ & $2 \, \Box$ \\  \hline \hline
	\end{tabular}
	\end{center}
	\caption{\label{Tab.1} Summary of the matrix elements appearing in $\delta^2\Gamma_k$ when expanded $\Gamma_k$ around flat Euclidean space. The column ``index'' identifies the corresponding matrix element in field space, $\Delta \equiv - \sib^{ij} \p_i \p_j$ is the Laplacian on the spatial slice, and $\Box \equiv -\p_t^2 - \sib^{ij} \p_i \p_j$. For each ``off-diagonal'' entry there is a second contribution involving the adjoint of the differential operator and the order of the fields reversed.}
\end{table}
On this basis, one can make the crucial observation that the component fields do not possess a relativistic dispersion relation. One may then attempt to add a suitable gauge-fixing term $\Gamma_k^{\rm gf}$. A suggestive choice (also from the perspective of Ho\v{r}ava-Lifshitz gravity) is proper-time gauge \cite{Dasgupta:2001ue}. This gauge choice eliminates the fluctuations in the lapse and shift vector $\Nh = 0$, $\Nh_i = 0$ by choosing 
\be
\Gamma_k^{\rm gf, proper-time} = \lim_{\alpha \rightarrow 0} \, \frac{1}{2\alpha} \, \int d\tau d^dy \sqrt{\sib} \, \left[ \Nh^2 + \Nh_i \, \sib^{ij} \, \Nh_j \right] \, . 
\ee
At the level of the component fields \eqref{TTshift} this choice entails $\Nh = 0$, $u_i = 0$ and $B=0$. This eliminates the last six entries from Tab.\ \ref{Tab.1}, essentially restricting quantum fluctuations to the components of the spatial metric. Tab.\ \ref{Tab.1} then indicates that the sector containing the fluctuations of the spatial metric (first five entries) contains propagators which do not include a spatial momentum dependence. On this basis 
proper-time gauge may not ideal for investigating the quantum properties of the theory in an off-shell formalism like the FRGE.
 
Motivated by the recent investigation \cite{Barvinsky:2015kil} it is then natural to investigate if there is a different gauge choice ameliorating this peculiar feature. Inspired by the decomposition \eqref{vdec} the gauge-fixing of the symmetries \eqref{eq:gaugeVariations} may be implemented via two functions $F$ and $F_i$  
\be\label{gf:ansatz}
\Gamma_k^{\rm gf} = \frac{1}{32 \pi G_k} \int d\tau d^dy \, \sqrt{\sib} \, \left[ F_i \, \sib^{ij} F_j + F^2 \right] \, , 
\ee
where $F$ and $F_i$ are linear in the fluctuation fields. The integrand entering $\Gamma^{\rm gf}_k$ may also be written in terms of a $D$-dimensional vector $F_\mu \equiv (F, F_i)$ and the background metric \eqref{FRWback} exploiting that $F_\mu \, \gb^{\mu\nu} \, F_\nu = F^2 + F_i \, \sib^{ij} \, F_j$. The most general form of $F$ and $F_i$ which is linear in the fluctuation fields $\Nh, \Nh_i, \sh_{ij}$ and involves at most one derivative with respect to the spatial or time coordinate is given by
\be\label{gauge1}
\begin{split}
	F = & \, c_1 \, \p_\tau \, \Nh + c_2 \, \p^i \, \Nh_i + c_3 \, \p_\tau \, \sh + d \, c_8 \, \Kb^{ij} \, \sh_{ij} + c_9 \, \Kb \Nh \, , \\
	F_i = & \, c_4 \, \p_\tau \, \Nh_i + c_5 \, \p_i \, \Nh + c_6 \, \p_i \, \sh + c_7 \, \p^j \, \sh_{ji} + d \, c_{10} \, \Kb_{ij} \Nh^j \, . 
\end{split}
\ee
The $c_i$ are real coefficients which may depend on $d$ and the factors $d$ are introduced for later convenience. Following the calculation in App.\ \ref{App.B3}, rewriting the gauge-fixing \eqref{gf:ansatz} in terms of the component fields yields \eqref{d2I5} and \eqref{d2I6}. Combining $\delta^2\Gamma_k^{\rm grav}$ with the gauge-fixing contribution one finally arrives at \eqref{eqS1B}. The coefficients $c_i$ are then fixed by requiring, firstly, that \emph{all component fields come with a relativistic dispersion relation} and, secondly, that the resulting gauge-fixed Hessian does not contain square-roots of the spatial Laplacian $\sqrt{\Delta}$. It turns out that these two conditions essentially fix the gauge uniquely, up to a physically irrelevant discrete symmetry:
\be\label{gffinal}
\begin{array}{lllll}
c_1= \epsilon_1 \, , \qquad & c_2= \epsilon_1 \, , \qquad & c_3=- \tfrac{1}{2} \epsilon_1 \, , \qquad & c_8=0 \, , \qquad & c_9= \tfrac{2\,(d-1)}{d} \,\epsilon_1 \, , \\[1.1ex]
	 c_4= \epsilon_2 \, \qquad & c_5=- \epsilon_2 \, , \qquad & c_6= - \half \epsilon_2 \, , \qquad & c_7= \epsilon_2  \, , \qquad & c_{10}= \tfrac{d-2}{d} \, \epsilon_2
\end{array} 
\ee
where $\epsilon_1=\pm 1$ and $\epsilon_2=\pm 1$. Since $\Gamma^{\rm gf}_k$ is quadratic in $F$ and $F_i$ it depends on $\epsilon_i^2$ only and the choice of sign does not change $\Gamma^{\rm gf}_k$.

Combining \eqref{GammaEH2} with the gauge choice \eqref{gf:ansatz} with \eqref{gffinal} finally results in the gauge-fixed Hessian
\be\label{eqS2Frank}
\begin{split}
	& \, 32  \pi G_k  \Big( \half \delta^2 \Gamma^{\rm grav}_k + \Gamma_k^{\rm gf} \Big) = \\ & \, 
	\int_x \Big\{ 
	\half \, h^{ij} \left[ \Delta_2 - 2 \Lambda_k - \tfrac{2(d-1)}{d} \dot{\Kb}
	- \tfrac{d^2-d+2}{d^2} \Kb^2 \right] \, h_{ij} \\ 	&  \qquad
	+ u^i \left[\Delta_1 -\tfrac{d-1}{d} \dot{\Kb} - \tfrac{1}{d} \Kb^2 \right] u_i 
	+ v^i \left[ \Delta_1 - 2 \Lambda_k  - \dot{\Kb} - \tfrac{5d-7}{d^2} \Kb^2  \right] v_i \\ &  \qquad
	+ B \, \left[ \Delta_0 - \tfrac{d-1}{d}\,\dot{\Kb}-\tfrac{d-1}{d^2}\Kb^2 \right] B 
	+ \Nh \left[ \Delta_0 - \tfrac{2(d-1)}{d} \dot{\Kb} {-\tfrac{4(d-1)}{d^2}}\, \Kb^2  \right] \Nh \\ 	&  \qquad 
	+  \Nh\, \Big[
	\Delta_0 - 2\Lambda_k -  {
	\tfrac{5d^2-12d+16}{4d^2}} \,\Kb^2 
	\Big] \big( (d-1) \psi + E \big) \\ &  \qquad
	- \tfrac{(d-1)(d-3)}{4} \, \psi \, \Big[  \Delta_0 - 2\Lambda_k  - {\tfrac{2(d-1)}{d} } \,\dot{\Kb} - { \tfrac{d-1}{d}} \Kb^2   \Big] \psi
	\\ &  \qquad
	+ \tfrac{1}{4} \, E \, \Big[ \Delta_0 -2 \Lambda_k  - {\tfrac{2(d-1)}{d}} \,\dot{\Kb} - {\tfrac{d-1}{d}} \,\Kb^2
	\Big]  E \\ &  \qquad
	- \tfrac{1}{2} (d-1) \, \psi \Big[ \Delta_0 - 2 \Lambda_k - {  \tfrac{2(d-1)}{d} \dot{\Kb} } - {\tfrac{d-1}{d}} \,\Kb^2
	\Big] E
	\Big\} \, . 
\end{split} 
\ee
Here the operators $\Delta_i$ are defined in \eqref{LapDala} and the diagonal terms in field space have been simplified by partial integration. Setting $\Kb = 0$, the matrix elements resulting from this expression are shown in the third column of Tab.\ \ref{Tab.1}. On this basis, it is then straightforward to verify that all fluctuation fields acquire a relativistic dispersion relation.
This condition fixes the gauge-choice \emph{uniquely} \cite{Biemans:2016rvp}.

The ghost action exponentiating the Faddeev-Popov determinant is obtained from the variations \eqref{eq:gaugeVariations} by evaluating \eqref{ghs}. The ghost sector then comprises one scalar ghost $\bar{c}, c$ and one spatial vector ghost $\bar{b}^i, b_i$ arising from the transformation of $F$ and $F_i$, respectively. Restricting to terms quadratic in the fluctuation field and choosing $\epsilon_1 = \epsilon_2 = -1$, the result is given by
\be\label{Gghost}
\begin{split}
\Gamma_k^{\rm ghost} = \int d\tau d^dy \sqrt{\sib} \, \Big\{ & {\bar{c} \left[\Delta_0 + \tfrac{2}{d} \Kb \p_\tau + \dot{\Kb}  \right] c } \\ & \, 
+ \bar{b}^i \left[ \Delta_1 + \tfrac{2}{d} \Kb \p_\tau + \tfrac{1}{d} \dot{\Kb} + \tfrac{d-4}{d^2} \Kb^2 \right] b_i \Big\} \, .
\end{split}
\ee
Notably, the ghost action does not contain a scale-dependent coupling. The results \eqref{eqS2Frank} and \eqref{Gghost} then complete the construction of the Hessian $\Gamma_k^{(2)}$.

At this stage the following remark is in order. Projectable Ho\v{r}ava-Lifshitz gravity \cite{Horava:2009uw} restricts the lapse function $N(\tau, y) \rightarrow N(\tau)$ to a function of time only while the symmetry group is restricted to foliation preserving diffeomorphisms $f(\tau,y) \rightarrow f(\tau)$. This structure suggests a Landau-type gauge-fixing for the lapse-function, setting $F = \Nh$. Retaining the most general (local) form of $F_i$ given in \eqref{gauge1}, a quick inspection of eq.\ \eqref{eqS1B} with $\Nh = 0$ and $c_1 = c_2=c_3 = 0$ reveals that there is no set of parameters $c_i$ which would bring the dispersion relations of the remaining component fields into the relativistic form displayed in Table \ref{Tab.1}. Thus the extension of the present off-shell construction to Ho\v{r}ava-Lifshitz gravity is not straightforward.

\subsection{Evaluating the operator traces}
\label{sect.3b}
Notably, the Hessians arising from \eqref{eqS2Frank} and \eqref{Gghost} contain $D$-covariant Laplace-type operators only and can thus be evaluated using standard heat-kernel techniques (see the Appendix of \cite{Codello:2008vh} for details). Resorting to a Type I regulator \cite{Codello:2008vh}, implicitly defined by 
\be
\Delta_s \mapsto P_k = \Delta_s + R_k \, , 
\ee
and choosing the profile function $R_k$ providing the $k$-dependent mass term for the fluctuation modes, to be of Litim-form $R_k=(k^2-\Delta_i)\,\theta(k^2-\Delta_i)$, the computation uses the  
heat-kernel techniques detailed in App.\ \ref{App.A}. Combining the intermediate results obtained in App.\ \ref{App.C},   
the flow of Newton's constant and the cosmological constant
is conveniently expressed in terms of 
%
%
 the dimensionless quantities
\be\label{defdimless}
\eta \equiv (G_k)^{-1} \p_t \, G_k \, , \qquad 
\lambda_k \equiv \Lambda_k \, k^{-2} \, , \qquad
g_k \equiv G_k \, k^{d-1} \, . 
\ee
Here $\eta$ is the anomalous dimension of Newton's constant. 
%
In order to write down the beta functions in a compact form, 
 it is moreover useful to define
\be
B_{\rm det}(\lambda) \equiv (1-2\lambda)(d-1-d\lambda)\, .
\ee
%
The final form of the beta functions is\footnote{The beta functions given here differ from the ones used in \cite{Biemans:2016rvp} by a different form of the regulator in the transverse-traceless and vector sectors of the decomposition \eqref{TTshift} and \eqref{TTmet}.}
\be\label{betafunction}
\begin{split}
	\beta_g = & \, (d-1+\eta) \, g \, , \\
	\beta_\lambda = & \, (\eta - 2) \lambda   + \tfrac{2g}{(4 \pi)^{(d-1)/2}} \, \tfrac{1}{\Gamma((d+3)/2)} 
	\Big[ 
	\big( d + 
	\tfrac{d^2 + d -4}{2(1-2\lambda)}
	+ \tfrac{3d-3 - (4 d-2) \lambda }{B_{\rm det}(\lambda)}
	\big) 
	\big( 1 - \tfrac{\eta}{d+3}  \big)  \, 
	\\ & \, \qquad \qquad  
	-2(d+1) + N_S + (d-1) N_V - 2^{\left[(d+1)/2\right] } \, N_D  \Big] \, , 
\end{split}
\ee
with anomalous dimension of Newton's constant  given by
\be\label{etaflow}
\begin{split}
	\eta = \frac{16 \pi g \, B_1(\lambda)}{(4\pi)^{(d+1)/2} + 16 \pi g \, B_2(\lambda)} \, : 
\end{split}
\ee
The functions $B_1(\lambda)$ and $B_2(\lambda)$ depend on $\lambda$ and $d$ and are given by
\be
\begin{split}
	 B_1(\lambda)   \equiv &
	- \tfrac{d^5 + 17 d^4 + 41 d^3 +  85 d^2 + 174 d - 78}{24 \, d ( d - 1) \, \Gamma(( d+5)/2)}
	+ \tfrac{d^4-5d^2+16d+48}{12 \, d (d-1)\, (1-2\lambda) \, \Gamma((d+1)/2)}
	\\ & \, 
	- \tfrac{d^4-15d^2+28d-10}{2 d (d-1) \, (1-2\lambda)^2 \, \Gamma((d+3)/2)} 
	%
	+ \tfrac{3d-3 - (4d-2) \lambda }{6 \, B_{\rm det}(\lambda) \,  \Gamma(( d+1)/2)} 
	+ \tfrac{c_{1,0} + c_{1,1} \lambda + c_{1,2} \lambda^2}{4 \, d \,   B_{\rm det}(\lambda)^2 \,  \Gamma((d+3)/2)} \\ & \, + \tfrac{1}{6 \, \Gamma((d+1)/2)} \left[N_S + \tfrac{d^2 - 13}{d+1} N_V  - \tfrac{1}{4} \, 2^{\left[ (d+1)/2\right]} N_D \, \right] \, . 
\end{split}
\ee
and
\be
\begin{split}
  	B_2(\lambda) = &  \, 
	\tfrac{d^4-10d^3+21d^2+6d+6}{24 \, d (d-1)\, \Gamma((d+5)/2)}
	+ \tfrac{d^4-5d^2+16d+48}{24 \,  d (d-1)\, (1-2\lambda) \, \Gamma((d+3)/2)}  
	- \tfrac{d^4-15d^2+28d-10}{4 \,d(d-1) \, (1-2\lambda)^2 \, \Gamma((d+5)/2)} \\ & \, 
	 + \tfrac{3d-3 - (4d-2)\lambda}{12 \, B_{\rm det}(\lambda) \, \Gamma((d+3)/2)}  
	+ \tfrac{c_{2,0} + c_{2,1} \lambda + c_{2,2} \lambda^2 }{8 \, d  \, B_{\rm det}(\lambda)^2 \,  \Gamma((d+5)/2)} \, . 
\end{split}
\ee
The coefficients $c_{i,j}$ are polynomials in $d$ and given by
\be
\begin{array}{ll}
c_{1,0} = - 5 d^3 + 22 d^2 - 24 d + 16 \, ,  \qquad &
c_{1,1} =  4 \left(d^3 - 10 d^2 + 16 d  -16 \right)
 \, ,  \\
c_{1,2} =  4  \left( d^3 + 6 d^2 - 16 d + 16  \right)  \, ,	&	
	c_{2,0} =  - 5 d^3 + 22 d^2 -24 d + 16      \, ,  
\end{array}
\ee
together with $c_{1,1} = c_{2,1}$ and $c_{1,2} = c_{2,2}$.
Notably $B_2$ is independent of the matter content of the system, reflecting the fact that the matter sector \eqref{matter} is independent of Newton's constant. The result \eqref{betafunction} together with the explicit expression for the anomalous dimension of Newton's constant \eqref{etaflow} constitutes the main result of this section.
     
\section{Properties of the RG flow}
\label{sect.4}
In this section, we analyze the RG flow resulting from the beta functions 
\eqref{betafunction} for a $D=3+1$-dimensional spacetime. The case of pure gravity, corresponding to setting $N_S = N_V = N_D = 0$, is discussed in Sect.\ \ref{sect.42} while the classification of the fixed point structure appearing in general gravity-matter systems is carried out in Sect.\ \ref{sect.43}. Our results complement the findings reported in \cite{Biemans:2016rvp}.

\subsection{Pure gravity}
\label{sect.42}
The beta functions \eqref{betafunction} constitute a system of coupled first-order differential equations. In general such systems do not admit analytical solutions and one has to resort to numerical methods. Nevertheless, the general theory of dynamical systems allows to determine possible long-term behaviors of the flow \eqref{betafunction} by determining its fixed points (FPs) $(g_*,\lambda_*)$ satisfying
\be\label{fpcond}
\beta_g(g_*,\lambda_*)=0\; , \qquad \beta_\lambda(g_*,\lambda_*)=0 \, . 
\ee
Such fixed points may control the long-term behavior of the theory in the limit $k\to\infty$ (UV completion) or $k\to0$ (IR limit). By linearizing the system \eqref{betafunction} around its FPs, the stability matrix $\mathbf{B}_{ij} \equiv \left. \p_{g_j} \beta_{g_i} \right|_{g = g_*}$ encodes the $k$-dependence of the couplings near the fixed point. In particular, the scaling of the couplings is characterized by the critical exponents $\theta_i$, defined by (minus) the eigenvalues of $\mathbf{B}_{ij}$:
eigendirections coming with ${\rm Re}(\theta_i) > 0$ are dragged into the fixed point for $k\to\infty$ while directions with ${\rm Re}(\theta_i) < 0$ are repelled in this limit. The former then constitute the relevant directions of the fixed point.

For $d=3$ spatial dimensions the system \eqref{betafunction} possesses a unique NGFP with positive Newton's constant,
\be\label{NGFP1}
\mbox{NGFP:} \qquad \; g_* = 0.785 \, , \qquad \lambda_* =  0.315 \, , \qquad g_* \lambda_* = 0.248 \, , 
\ee
coming with a complex pair of critical exponents, 
\be\label{critexp}
\theta_{1,2} = 0.503 \pm  5.377 i \, . 
\ee
The positive real part, Re($\theta_{1,2}) > 0$, indicates that the NGFP  acts as a spiraling UV attractor for the RG trajectories in its vicinity. Notably, this is the same type of UV-attractive spiraling behavior encountered when evaluating the RG flow on foliated spacetimes using the Matsubara formalism \cite{Manrique:2011jc,Rechenberger:2012dt}, and a vast range of studies building on the metric formalism \cite{Souma:1999at,Lauscher:2001ya,Reuter:2001ag,Litim:2003vp,Fischer:2006fz,Codello:2013fpa,Nagy:2013hka,Christiansen:2015rva,Codello:2006in,Codello:2007bd,Benedetti:2009rx,Benedetti:2009gn,Demmel:2014hla,Machado:2007ea,Manrique:2010am,Christiansen:2012rx,Christiansen:2014raa,Falls:2014tra,Falls:2015cta,Donkin:2012ud,Lauscher:2002sq,Groh:2010ta,Manrique:2009uh,Manrique:2010mq,Becker:2014qya,Gies:2015tca,Eichhorn:2009ah,Rechenberger:2012pm,Eichhorn:2010tb,Nink:2012vd,Becker:2014pea,Becker:2014jua}.

\begin{figure}[t!]
	\begin{center}
		\includegraphics[width=0.48\textwidth]{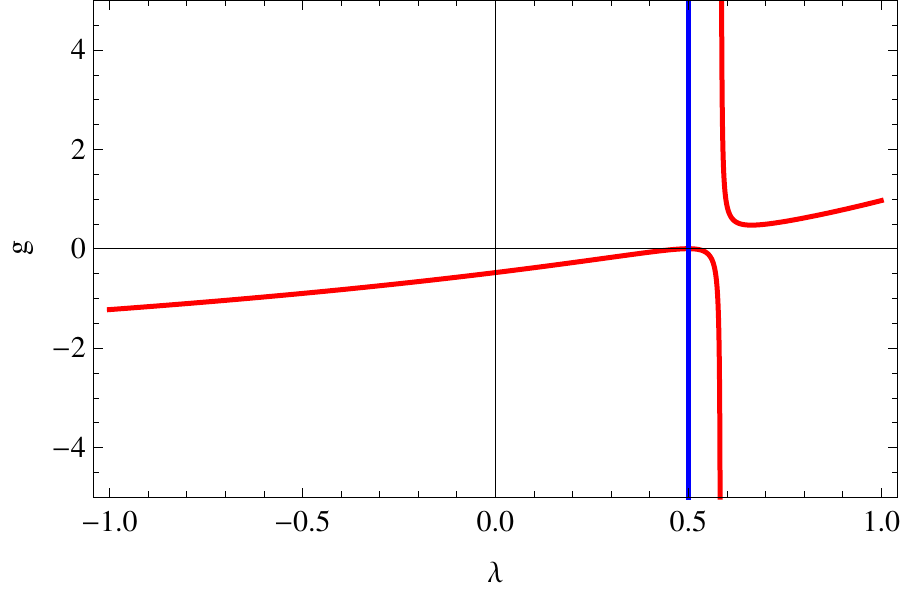} \; 
		\includegraphics[width=0.48\textwidth]{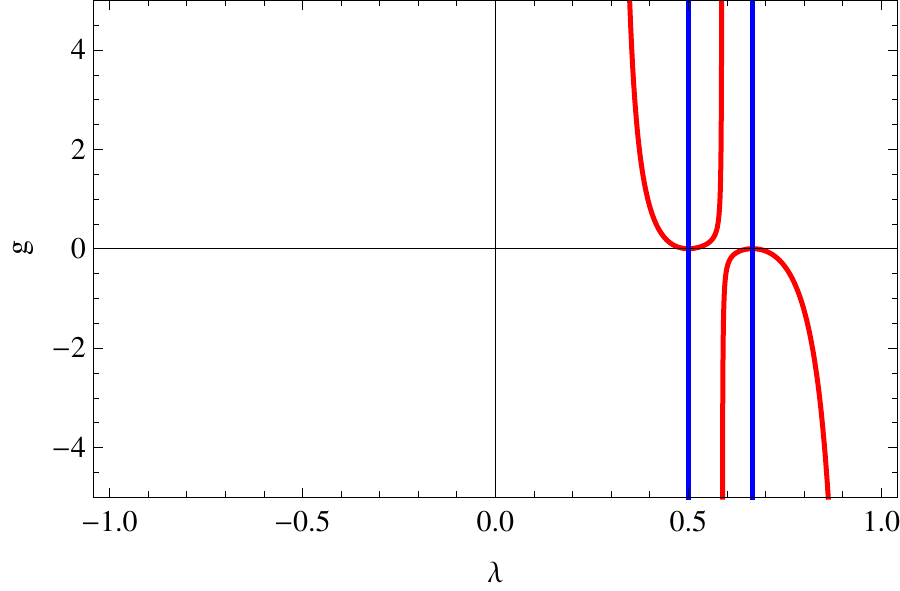} 
		\caption{Singularity structure of the beta functions \eqref{betafunction} in the $\lambda$-$g$--plane in $d=2$ (left diagram) and $d=3$ (right diagram). The blue lines indicate fixed singularities of $\beta_\lambda$, eq.\ \eqref{singlin1}, while the red lines illustrate the curves \eqref{etasing2} and where $\eta$ develops a singularity. \label{Fig.sing} } 
	\end{center}	 
\end{figure}

Subsequently, it is instructive to determine the singular loci of the beta functions \eqref{betafunction} where either $\beta_g$ or $\beta_\lambda$ diverge. For finite values of $g$ and $\lambda$ these may either be linked to  
one of the denominators appearing in $\beta_\lambda$ becoming zero or a divergences of the anomalous dimension of Newton's constant.
Inspecting $\beta_\lambda$, the first case gives rise to two singular lines in the $\lambda$-$g$--plane
\be\label{singlin1}
\begin{split}
	\lambda^{\rm sing}_1 = \tfrac{1}{2} \, , \qquad \lambda^{\rm sing}_2 = \tfrac{d-1}{d} \, . 
\end{split}
\ee
The singular lines $\eta^{\rm sing}(g,\lambda)$ associated with divergences of the anomalous dimension $\eta$ are complicated functions of $d$. For the specific cases $d=2$ and $d=3$ the resulting expressions simplify and are given by the parametric curves 
\be\label{etasing2}
\begin{split}
	\begin{array}{lll}
	d = 2: \; \qquad &	\eta^{\rm sing}: \qquad & g = - \frac{45 \pi (1-2\lambda)^2}{2 \, (76 \lambda^2 - 296 \lambda + 147 ) }    ,  \\[1.2ex] 
	d = 3: \; \qquad &	\eta^{\rm sing}: \qquad & g = - \frac{144 \pi (6 \lambda^2 - 7 \lambda + 2)^2}{144 \lambda^4 -1884 \lambda^3 +3122 \lambda^2 -1688 \lambda + 279 } \, .  
	\end{array}
\end{split}
\ee
The position of the singular lines \eqref{singlin1} and \eqref{etasing2} are illustrated in Fig.\ \ref{Fig.sing}. Focusing to the domain $g \ge 0$, it is interesting to note that the singularities bounding the flow of $\lambda_k$ for positive values are of different nature in $d=2$ and $d=3$: in $d=2$ the domain is bounded to the right by a fixed singularity of $\beta_\lambda$ and $\eta$ remains finite throughout this domain while in $d=3$ the singular line $\lambda_1^{\rm sing}$ is screened by a divergence of $\eta$. Notably, the position of the singular lines is independent of $N_S$, $N_V$, and $N_D$ and thus also carries over to the analysis of gravity-matter systems.

Finally, we note that the point $(\lambda,g) = (1/2,0)$ is special in the sense that the beta functions \eqref{betafunction} are of the form $0/0$. In particular the value of the anomalous dimension $\eta$ depends on the direction along which this point is approached. We will denote this point as ``quasi-fixed point'' $C\equiv(\tfrac{1}{2},0)$ in the sequel. 

\begin{figure}[t!]
	\begin{center}
		\includegraphics[width=0.8\textwidth]{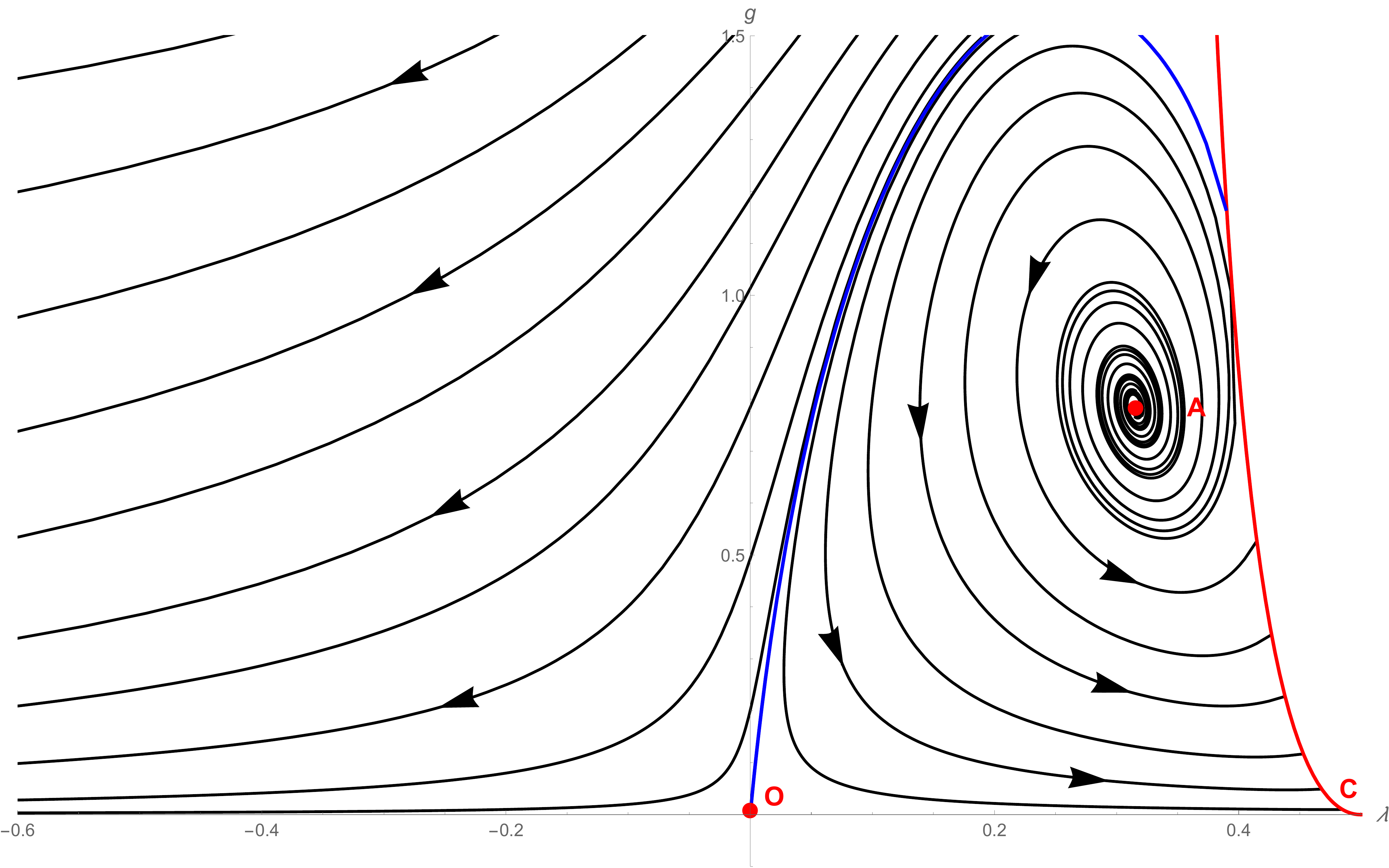}
		\caption{Phase diagram of the RG flow originating from the beta functions \eqref{betafunction} in $D=3+1$ spacetime dimensions. The flow is dominated by the interplay of the NGFP (point ``A'') controlling the flow for ultra-high energies and the GFP (point ``O'') governing the low-energy behavior. The flow undergoes a crossover between these two fixed points. For some of the RG trajectory this crossover is intersected by the singular locus \eqref{etasing2} (red line). The arrows indicate the direction of the RG flow pointing from high to low energy.
			\label{flow2d3d} } 
	\end{center}	 
\end{figure}

Upon determining the fixed point and singularity structure relevant for the renormalization group flow with a positive Newton's constant, it is rather straightforward to construct the RG trajectories resulting from the beta functions \eqref{betafunction} numerically. An illustrative sample of RG trajectories characterizing the flow in $D=3+1$ spacetime dimensions is shown in Fig.\ \ref{flow2d3d}.
Notably, the high-energy behavior of the flow is controlled by
the NGFP \eqref{NGFP1}. Following the nomenclature introduced in \cite{Reuter:2001ag}, the low-energy behavior can be classified according to the sign of the cosmological constant:
\be\label{Typeclass}
\begin{array}{lll}
\mbox{Type Ia:}  \qquad \quad & \lim_{k \rightarrow 0} (\lambda_k, g_k) = (- \infty, 0) \, , \qquad   & \Lambda_0 < 0 \, ,  \\[1.1ex]
\mbox{Type IIa:}  & \lim_{k \rightarrow 0} (\lambda_k, g_k) = (0, 0) \, ,  & \Lambda_0 = 0  \, , \\[1.1ex]
\mbox{Type IIIa:} & \mbox{terminate at} \; \; \eta^{\rm sing} & \Lambda_{k_{\rm term}} > 0 \, . \\
\end{array}
\ee
These three phases are realized by the RG trajectories flowing to the left (Type Ia), to the right (Type IIIa), and on top of the bold blue line emanating from the GFP ``O'' (Type IIa). Once the trajectories enter the vicinity of the GFP, characterized by $g_k \ll 1$, the dimensionful Newton's constant $G_k$ and cosmological constant $\Lambda_k$ are essentially $k$-independent, so that the trajectories enter into a ``classical regime''. For trajectories of Type Ia this regime extends to $k = 0$. Trajectories of Type IIIa terminate in the singularity $\eta^{\rm sing}$ (red line) at a finite value $k_{\rm term}$.
The high-energy and low-energy regimes are connected by a crossover of the RG flow. For some of the trajectories, this crossover cuts through the red line marking a divergence in the anomalous dimension of Newton's constant.
This peculiar feature can be traced back to the critical exponents \eqref{critexp} where the beta functions \eqref{betafunction} lead to a exceptionally low value for Re($\theta_{1,2}$). Compared to other incarnations of the flow, which come with significantly higher values for Re($\theta_{1,2}$), this makes the spiraling process around the NGFP less compact. As a consequence the flow actually touches $\eta^{\rm sing}$. Since this feature is absent in the flow diagrams obtained from the Matsubara computation \cite{Manrique:2011jc,Rechenberger:2012dt}, the foliated RG flows studied in \cite{Biemans:2016rvp}, and in the flows obtained in the covariant formalism \cite{Reuter:2001ag}, it is likely that this is rather a particularity of the flow based on \eqref{betafunction}, instead of a genuine physical feature.

\subsection{Gravity-matter systems}
\label{sect.43}
%
In order to classify the fixed point structures realized for a generic gravity-matter system, we first observe that the number of minimally coupled 
scalar fields, $N_S$, vectors, $N_V$, and Dirac spinors $N_D$ enter the gravitational beta functions \eqref{betafunction} in terms of the combinations\footnote{The precise relation between the parameters $d_g$, $d_\lambda$ and the matter content may depend on the precise choice of regulator employed in matter traces, see App.\ \ref{App.C3}. Carrying out the classification of fixed point structures in terms of the deformation parameters shifts this regulator dependence into the map $d_g(N_S,N_V,N_D), d_\lambda(N_S,N_V,N_D)$ allowing to carry out the classification independently of a particular regularization scheme.} 
\be
d_g \equiv N_S + \frac{d^2-13}{d+1} N_V - \tfrac{1}{4} \, 2^{(d+1)/2} N_D  \, , \quad
d_\lambda \equiv N_S + (d-1) N_V - 2^{(d+1)/2} N_D 
 \, . 
\ee
For $d=3$ these definitions reduce to 
\be\label{dldgdef}
d_g = N_S - N_V - N_D  \, , \quad
d_\lambda = N_S + 2 N_V - 4 N_D
  \, . 
\ee
The relation \eqref{dldgdef} allows to assign coordinates to any matter sector. For example, the standard model of particle physics comprises $N_S =4$ scalars, $N_D = 45/2$ Dirac fermions and $N_V = 12$ vector fields and is thus located at $(d_g, d_\lambda) = (- 61/2, - 62)$. For $N_S$ and $N_V$ being positive integers including zero and $N_D$ taking half-integer values in order to also accommodate chiral fermions, $d_g$ and $d_\lambda$ take half-integer values and cover the entire $d_g$-$d_\lambda$--plane.  

The beta functions \eqref{betafunction} then give rise to a surprisingly rich set of NGFPs whose properties can partially be understood analytically. The condition $\beta_g|_{g = g_*} = 0$ entails that any NGFP has to come with an anomalous dimension $\eta_* = -2$. This relation can be solved analytically, determining the fixed point coordinate $g_*(\lambda_*; d_g)$ as a function of $\lambda_*$ and $d_g$. Substituting $\eta_* = -2$ together with the relation for $g_*$ into the second fixed point condition, $\beta_\lambda|_{g = g_*}  = 0$, then leads to a fifth order polynomial in $\lambda$ whose coefficients depend on $d_g, d_\lambda$. The roots of this polynomial provide the coordinate $\lambda_*$ of a candidate NGFP. The fact that the polynomial is of fifth order then entails that the beta functions \eqref{betafunction} may support at most five NGFPs, independent of the matter content of the system. 
\begin{figure}[t!]
	\begin{center}
		\includegraphics[width=0.7\textwidth]{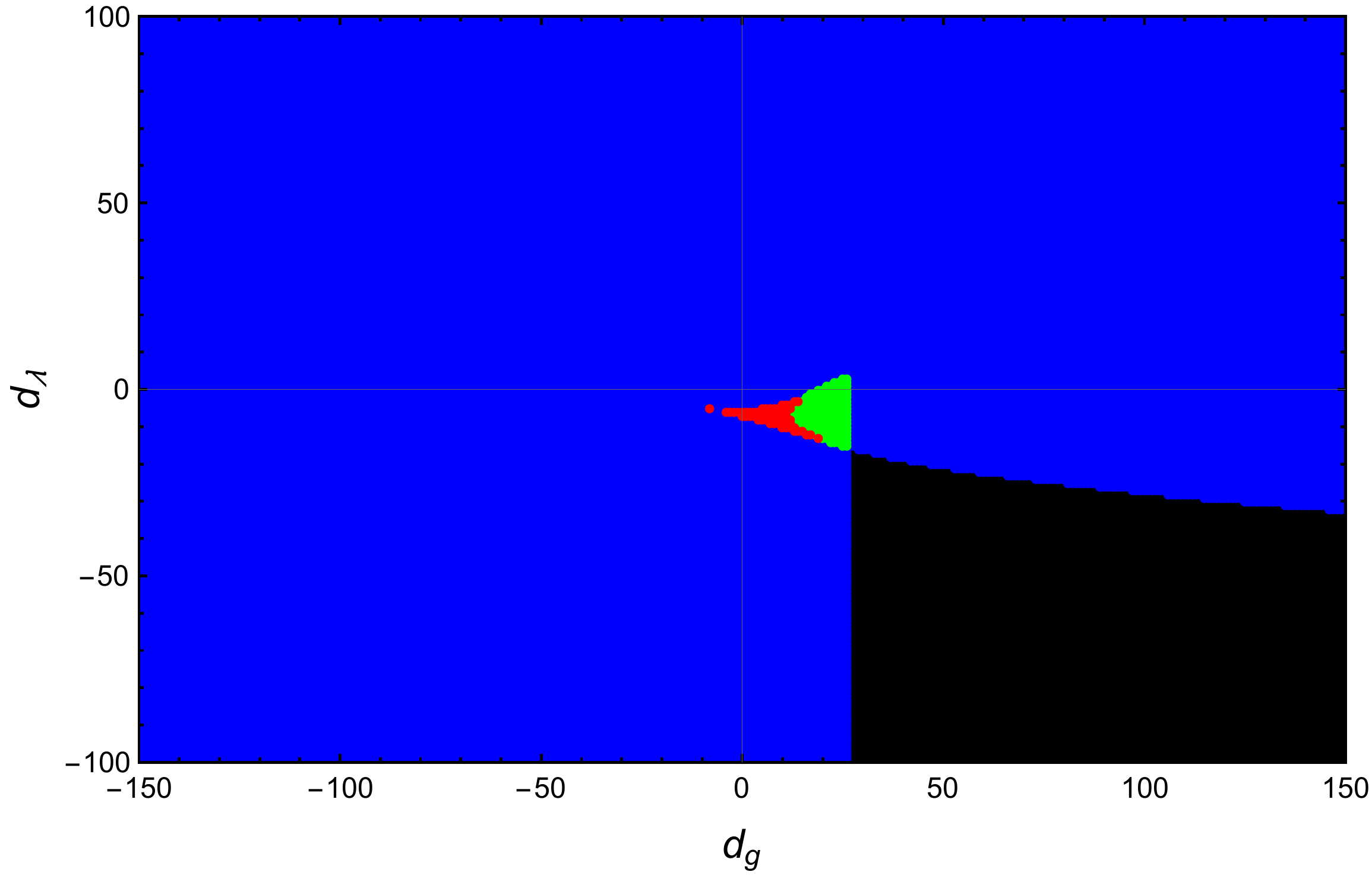}
		\caption{Number of NGFPs supported by the beta functions \eqref{betafunction} as a function of the parameters $d_g$ and $d_\lambda$. The colors black, blue, green, and red indicate the existence of zero, one, two, and three NGFPs situated at $g_* > 0$, $\lambda_* < 1/2$, respectively. \label{Fig.7a}} 
	\end{center}	 
\end{figure}

\begin{figure}[t!]
	\begin{center}
		\includegraphics[width=0.489\textwidth]{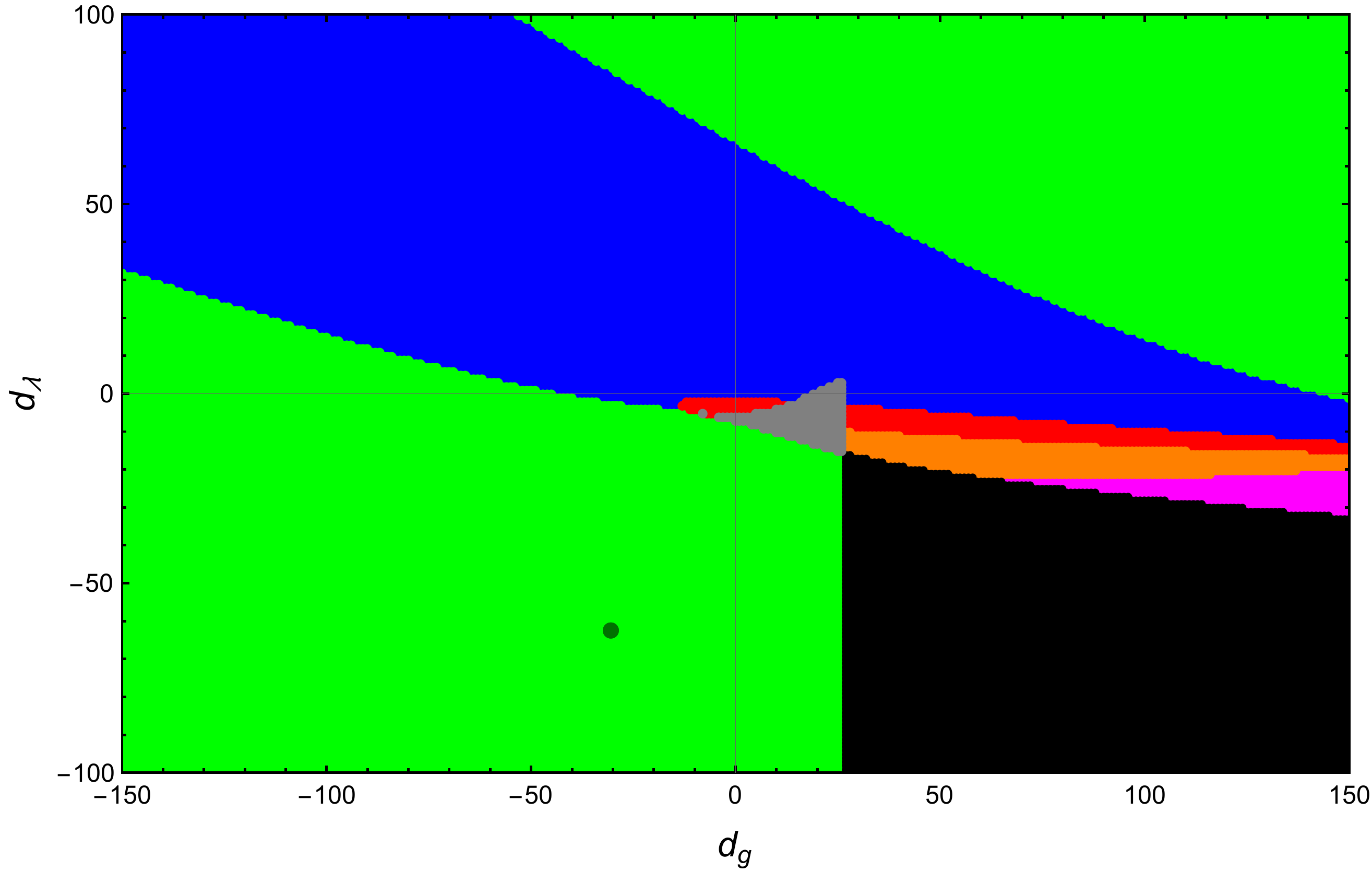} \; 
		\includegraphics[width=0.487\textwidth]{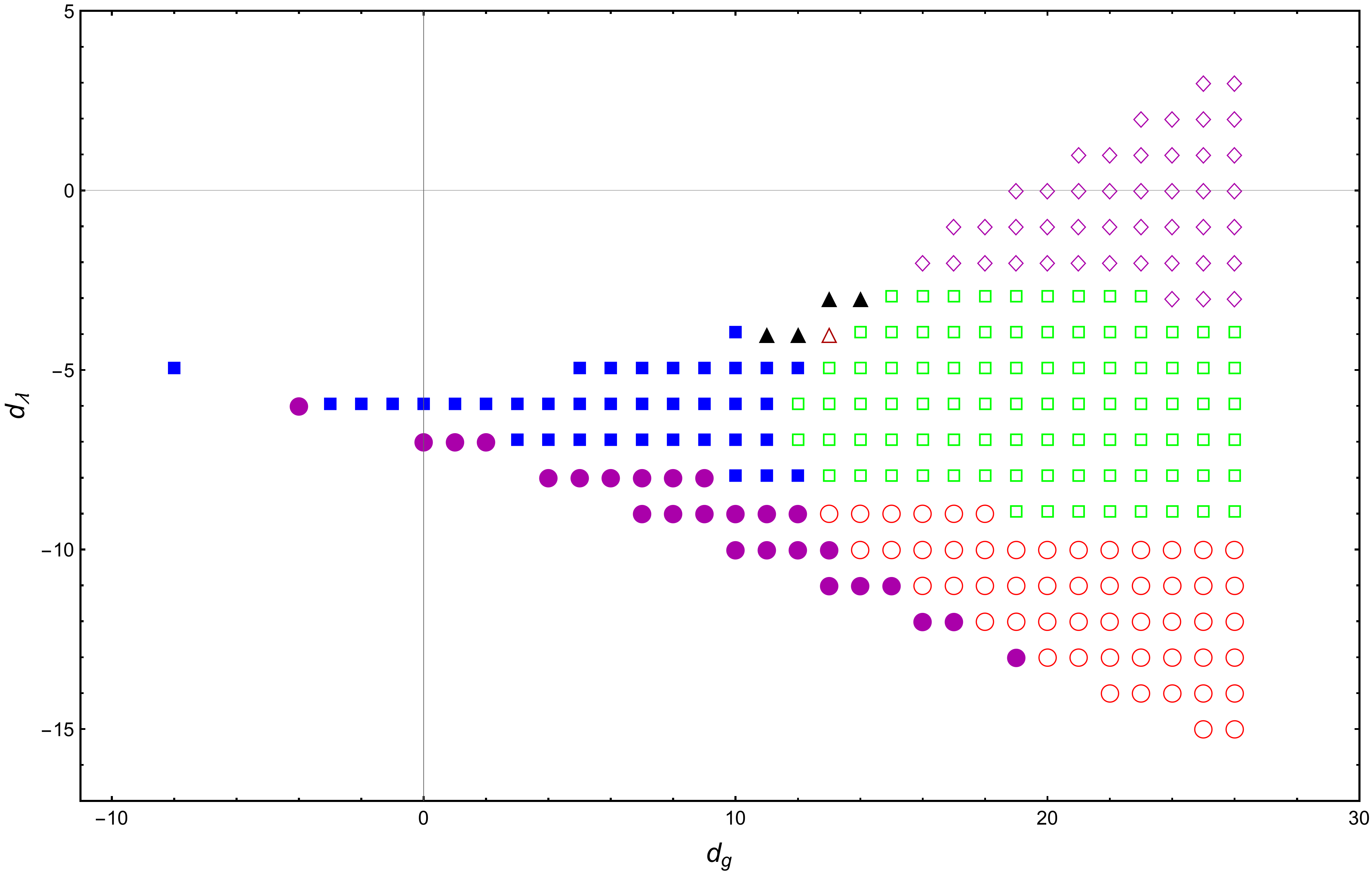} 
		\caption{\label{Fig.7} Classification of the NGFPs arising from the beta functions \eqref{betafunction} in the $d_g$-$d_\lambda$--plane, following the color-code provided in Table \ref{Tab.4}. The left diagram classifies the stability behavior of the one-fixed point sector. In particular, the black region does not support any NGFP while the regions giving rise to a single, UV-attractive NGFP with complex and real critical exponents are marked in blue and green, respectively. The field content of the standard model is situated in the lower-left quadrant, $(d_g, d_\lambda) = (-61/2, -62)$, and marked with a bold green dot. The gray area, supporting multiple NGFPs is magnified in the right diagram with empty and filled symbols indicating the existence of two and three NGFPs, respectively.} 
	\end{center}	 
\end{figure}

The precise fixed point structure realized for a particular set of values $(d_g, d_\lambda)$ can be determined numerically. The number of NGFPs located within the physically interesting region $g_* > 0$ and $\lambda_* < 1/2$ is displayed in Fig.\ \ref{Fig.7a}, where black, blue, green and red mark matter sectors giving rise to zero, one, two, and three NGFPs, respectively. On this basis, we learn that systems possessing zero or one NGFP are rather generic, while matter sectors giving rise to two or three NGFPs are confined to a small region in the center of the $d_g$-$d_\lambda$--plane.
\begin{table}[t!]
	\renewcommand{\arraystretch}{1.4}
	\begin{center}
		\begin{tabular}{|c||c||c|c|c||c|} \hline \hline
			class & NGFPs & NGFP$_1$ & NGFP$_2$ & NGFP$_3$ & color code \\ \hline \hline
			Class 0 & 0 & $-$ & $-$ & $-$ & black region  \\ \hline \hline
			Class Ia & 1 & UV, spiral & $-$ & $-$ & blue region \\ \hline
			Class Ib & 1 & UV, real & $-$ & $-$ & green region \\ \hline
			Class Ic & 1 & saddle & $-$ & $-$ & magenta region \\ \hline
			Class Id & 1 & IR, spiral & $-$ & $-$ & red region \\ \hline
			Class Ie & 1 & IR, real & $-$ & $-$ & orange region \\ \hline \hline
			Class IIa & 2 & UV, real & IR, real & $-$ & open circle \\ \hline
			Class IIb & 2 & UV, real & IR, spiral & $-$ & open square \\ \hline
			Class IIc & 2 & UV, spiral & IR, spiral & $-$ & open triangle \\ \hline
			Class IId & 2 & UV, spiral & UV, real & $-$ & open diamond \\ \hline
			\hline
			Class IIIa & 3 & UV, real & saddle & IR, real & filled circle \\ \hline
			Class IIIb & 3 & UV, real & saddle & IR, spiral & filled square \\ \hline
			Class IIIc & 3 & UV, spiral & saddle & IR, spiral & filled triangle \\ \hline
			\hline
		\end{tabular}
	\end{center}
	\caption{\label{Tab.4} Color-code for the fixed point classification provided in Fig.\ \ref{Fig.7}. The column NGFPs gives the number of NGFP solutions while the subsequent columns characterize their behavior in terms of 2 UV-attractive (UV), one UV-attractive and one UV-repulsive (saddle) and 2 IR-attractive (IR) eigendirections with real (real) and complex (spiral) critical exponents.}
\end{table}

The classification of the NGFPs identified in Fig.\ \ref{Fig.7a} according to their stability properties is provided in Fig.\ \ref{Fig.7} with the color-coding explained in Table \ref{Tab.4}. The left diagram provides the classification for the case of zero (black region) and one NGFP. Here 
 green and blue indicate the existence of a single UV-attractive NGFP with real (green) or complex (blue) critical exponents. Saddle points with one UV attractive and one UV repulsive eigendirection (magenta) and IR fixed points (red, orange) occur along a small wedge paralleling the $d_\lambda$-axis, only. The gray region supporting multiple NGFPs is magnified in the right diagram of Fig.\ \ref{Fig.7a}. All points in this region support at least one UV NGFP suitable for Asymptotic Safety while there is a wide range of possibilities for the stability properties of the second and third NGFP. Clearly, it would be interesting to study the RG flow resulting from the interplay of these fixed points. Since it will turn out (see Table \ref{Tab.3}), however, that there is no popular particle physics model situated in this regions, we postpone this investigation to a subsequent work.

The classification in Fig.\ \ref{Fig.7a} establishes that the existence of
a UV-attractive NGFP suitable for Asymptotic Safety is rather generic and puts only mild constraints on the admissible values $(d_g, d_\lambda)$. At this stage, it is interesting to relate this classification to phenomenologically interesting matter sectors including the standard model of particle physics (SM) and its most commonly studied extensions.\footnote{For a similar discussion within metric approach to Asymptotic Safety see \cite{Dona:2013qba}.} The result is summarized in Table \ref{Tab.3}. The map \eqref{dldgdef} allows to relate the number of scalars $N_S$, vector fields $N_V$ and Dirac fermions $N_D$ defining the field content of a specific matter sector to coordinates in the $d_g$-$d_\lambda$--plane. The resulting coordinates are given in the fifth and sixth column of Table \ref{Tab.3}. 
\begin{table}[t!]
	\renewcommand{\arraystretch}{1.4}
	\begin{center}
		\begin{tabular}{|p{3.94cm}||c|c|c||c|c||c|c||c|c|} \hline \hline
			model & $N_S$ & $N_D$ & $N_V$ & $d_g$ & $d_\lambda$ & $g_*$ & $\lambda_*$ & $\theta_1$ & $\theta_2$ \\  \hline \hline
			pure gravity & 0 & 0 & 0 &  0 & 0 & $0.78$ & $+\,0.32$ & \multicolumn{2}{c|}{$0.50 \pm 5.38 \, i$} \\ \hline
			Standard Model (SM) & 4 & $\tfrac{45}{2}$ & 12 & $-\,\tfrac{61}{2}$ & $-\,62$ & $0.75$ & $-\,0.93$ & 3.871 & 2.057 \\ \hline
			SM, dark matter (dm) & 5 & $\tfrac{45}{2}$ & 12 &  $-\,\tfrac{59}{2}$ & $-\,61$ & $0.76$ & $-\,0.94$ & 3.869 & 2.058 \\ \hline
			SM, $3\,\nu$ & 4 & 24 & 12 & $-\,32$ & $-\,68$ & $0.72$ & $-\,0.99$ & 3.884 & 2.057 \\ \hline
			SM, $3\,\nu$, dm, axion & 6 & 24 & 12 & $-\,30$ & $-\,66$ & $0.75$ & $-\,1.00$ & 3.882 & 2.059 \\ \hline
			MSSM & 49 & $\tfrac{61}{2}$ & 12 & $+\,\tfrac{13}{2}$ & $-\,49$ & $2.26$ & $-\,2.30$ & 3.911 & 2.154 \\ \hline
			{SU(5) GUT} & {124} & {24} & {24} &  $+\,76$ & $+\,76$ & $0.17$ & $+\,0.41$ & 25.26 & 6.008 \\ \hline
			{SO(10) GUT} & {97} & {24} & {45} & $+\,28$ & $+\,91$ & $0.15$ & $+\,0.40$ & 19.20 & 6.010 \\ \hline \hline
		\end{tabular}
	\end{center}
	\caption{\label{Tab.3} Fixed point structure arising from the field content of commonly studied matter models. All models apart from the minimally supersymmetric standard model (MSSM) and the grand unified theories (GUT), sit in the lower--left quadrant of Fig.\ \ref{Fig.7}. All matter configurations possess a single ultraviolet attractive NGFP with real critical exponents.}
\end{table}
Correlating these coordinates with the data provided by Fig.\ \ref{Fig.7} yields two important results: Firstly, all matter models studied in Table \ref{Tab.3} are located in regions of the $d_g$-$d_\lambda$--plane which host a single UV-attractive NGFP with real stability coefficients. Secondly, we note a qualitative difference between the standard model and its extensions (first five matter sectors) and grand unified theories (GUTs). The former all belong to the green region in the lower left part of the $d_g$-$d_\lambda$--plane while the second class of models sits in the upper-right quadrant. As a result, the corresponding NGFPs possess very distinct features. The NGFPs appearing in the first case have a characteristic product $g_* \lambda_* < 0$. Their critical exponents show a rather minor dependence on the precise matter content of the theory and have values in the range $\theta_1 \simeq 3.8 - 3.9$ and $\theta_2 \simeq 2.0$. In contrast, the NGFPs appearing in the context of GUT-type models come with a positive product $g_* \lambda_* > 0$. Their are significantly larger $\theta_1 > 19$ than in the former case and show a much stronger dependence on the matter field content. Thus while all matter sectors investigated in Table \ref{Tab.3} give rise to a NGFP suitable for realizing Asymptotic Safety  the magnitude of the critical exponents hints that the SM-type theories may have more predictive power in terms of a lower number of relevant coupling constants in the gravitational sector.   

At this stage it is also instructive to construct the phase diagram resulting from gravity coupled to the matter content of the standard model.
\begin{figure}[t!]
	\begin{center}
		\includegraphics[width=0.8\textwidth]{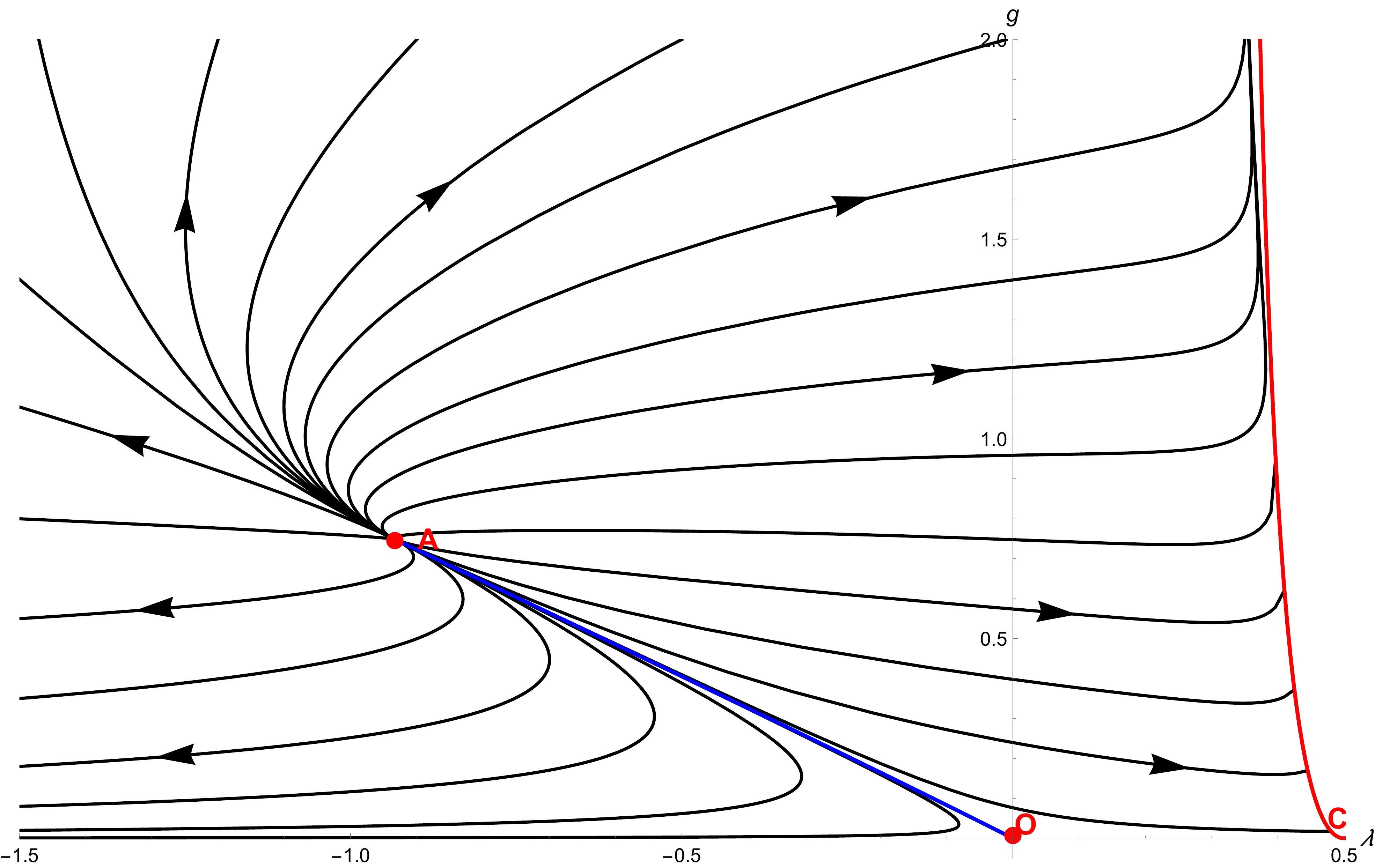}
		\caption{Phase diagram depicting the RG flow of gravity coupled to the matter content of the standard model in $D=3+1$ spacetime dimensions. Similarly to the pure gravity case, the phase diagram is dominated by the interplay of the NGFP (point ``A'') controlling the flow for ultra-high energies and the GFP (point ``O'') governing its low-energy behavior. The singular locus \eqref{etasing2} is depicted by the red line and arrows point towards lower values of $k$.
			\label{flowSM} } 
	\end{center}	 
\end{figure}
Following the strategy of Sect.\ \ref{sect.42}, an illustrative sample of RG trajectories obtained from solving the beta functions \eqref{betafunction} for $(d_g, d_\lambda) = (- 61/2, - 62)$ is shown in Fig.\ \ref{flowSM}. Similarly to the case of pure gravity, the flow is dominated by the interplay of the NGFP situated at $(g_*,\lambda_*) = (0.75,-0.93)$ and the GFP in the origin. The NGFP controls the UV behavior of the trajectories while the GFP is responsible for the occurrence of a classical low-energy regime. The classification of possible low-energy behaviors is again given by the limits \eqref{Typeclass}. A notable difference to the pure gravity case is the absence of the inspiraling behavior of trajectories onto the NGFP. This reflects the property that the NGFPs of the gravity-matter models come with real critical exponents. Moreover, the shift of the NGFP to negative values $\lambda_*$ entails that the singularity \eqref{etasing2} (red line) no longer affects the crossover of the trajectories from the NGFP to the GFP. Notably, other matter sectors located in the lower-left green region of Fig.\ \ref{Fig.7} give rise to qualitatively similar phase diagrams so that the flow shown in Fig.\ \ref{flowSM} provides a prototypical showcase for this class of universal behaviors.

Owed to their relevance for cosmological model building, we close this section with a more detailed investigation of the fixed point structures appearing in gravity-scalar models with $N_V = N_D = 0$. For illustrative purposes we formally also include negative values $N_S$ in order to capture the typical behavior of matter theories located in the lower-left quadrant of Fig.\ \ref{Fig.7}. 
\begin{figure}[t!]
	\begin{center}
		\includegraphics[width=0.45\textwidth]{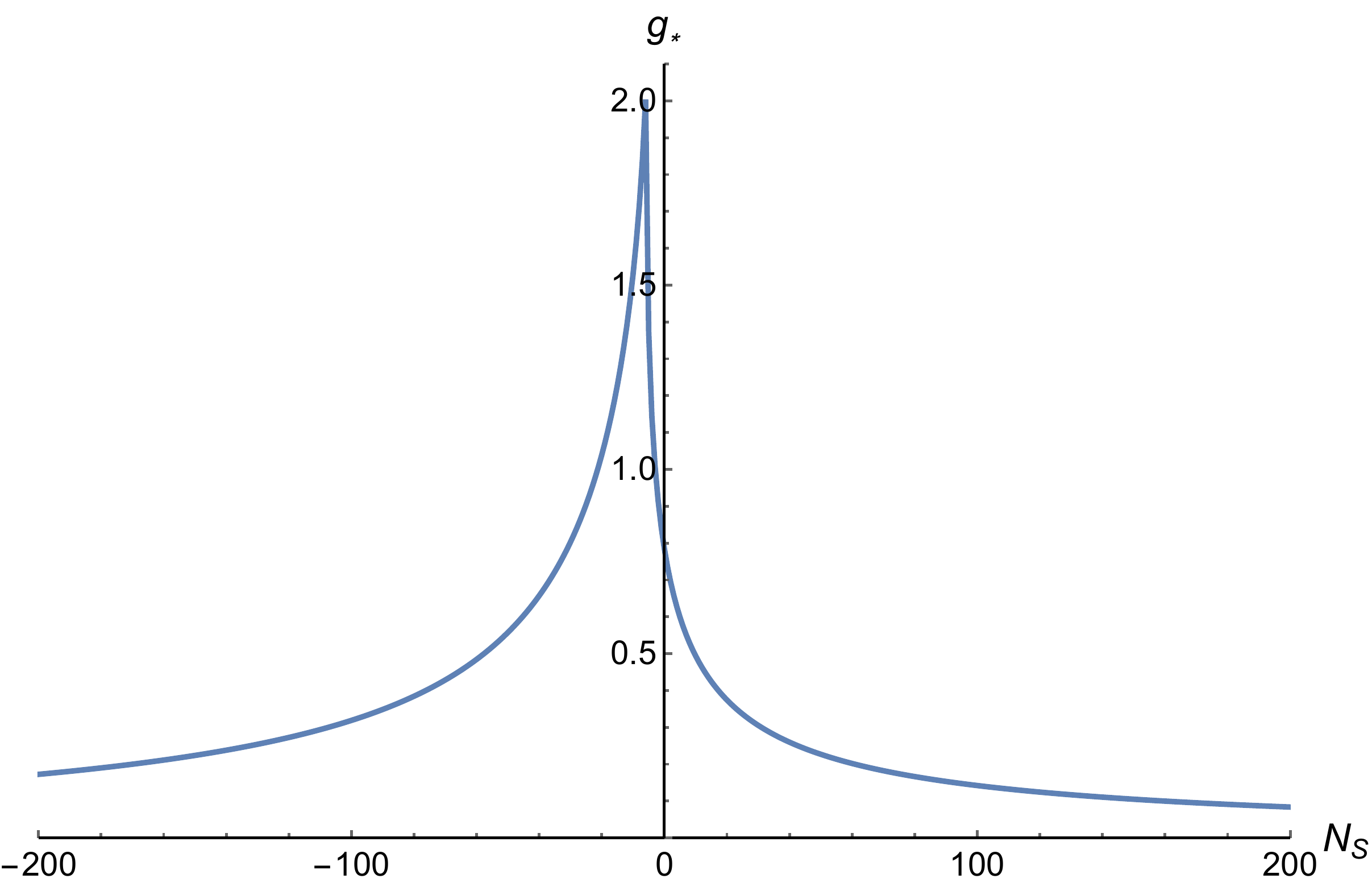} \; \;
		\includegraphics[width=0.45\textwidth]{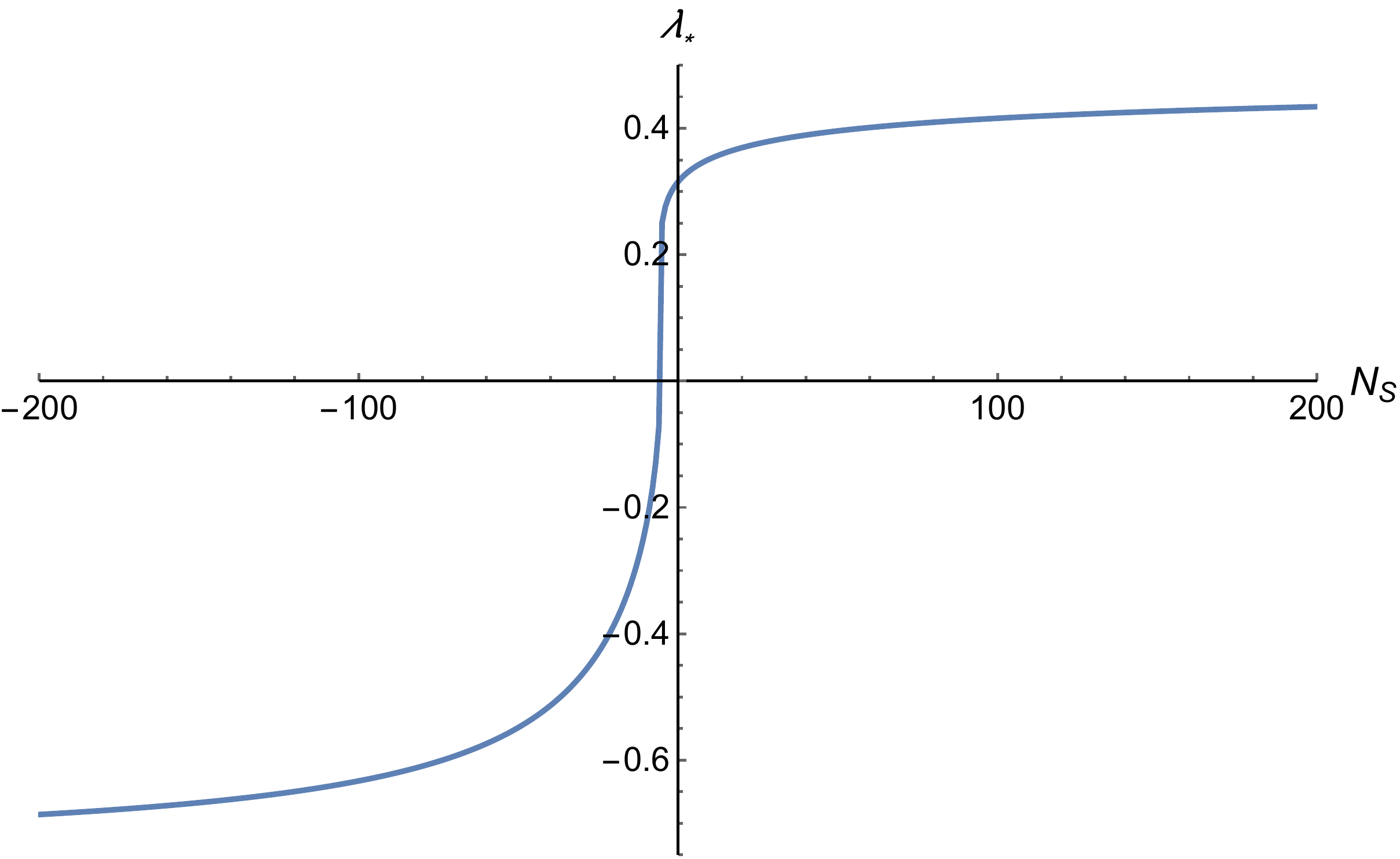} \\[2ex]
		\includegraphics[width=0.45\textwidth]{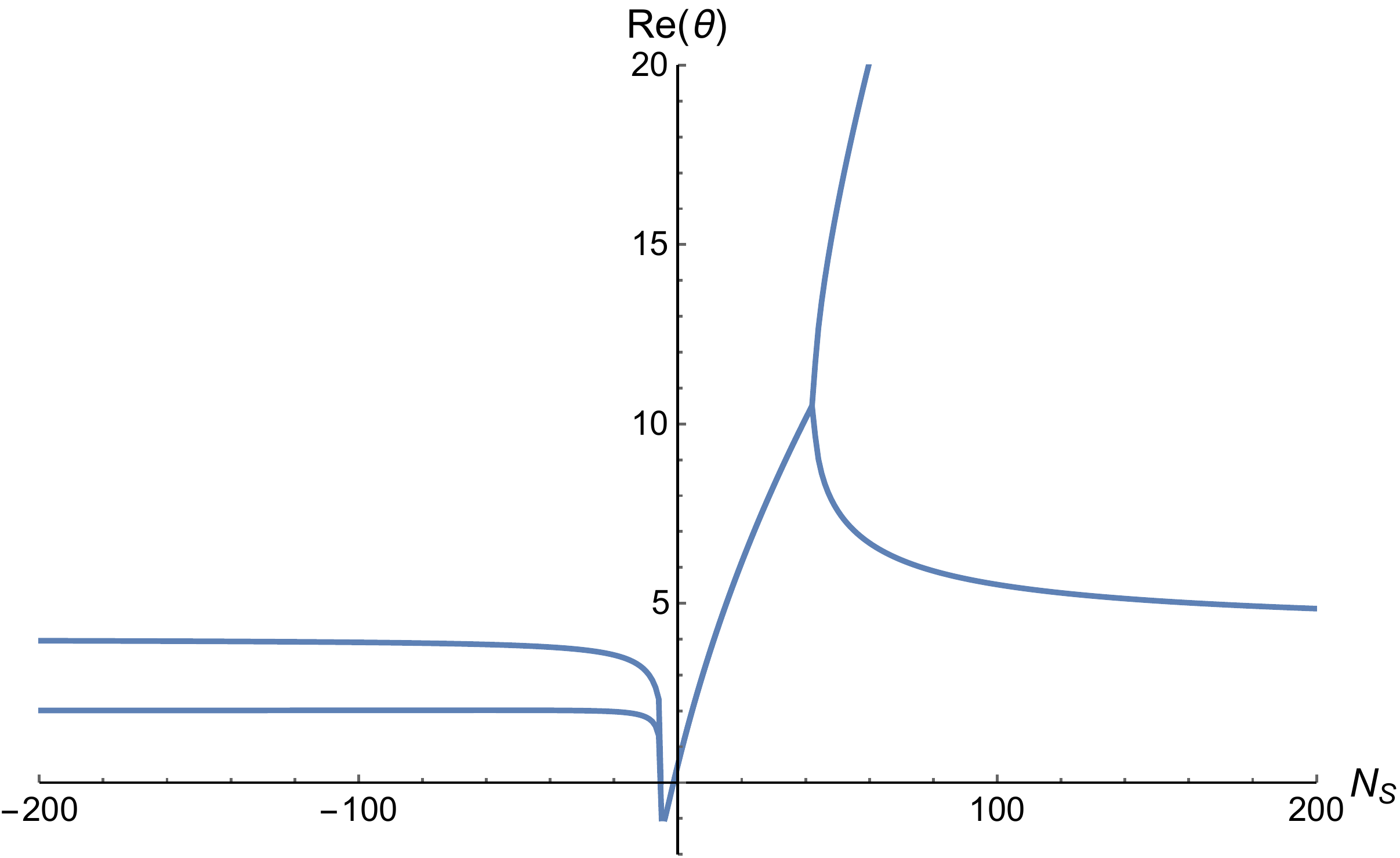} \; \;
		\includegraphics[width=0.45\textwidth]{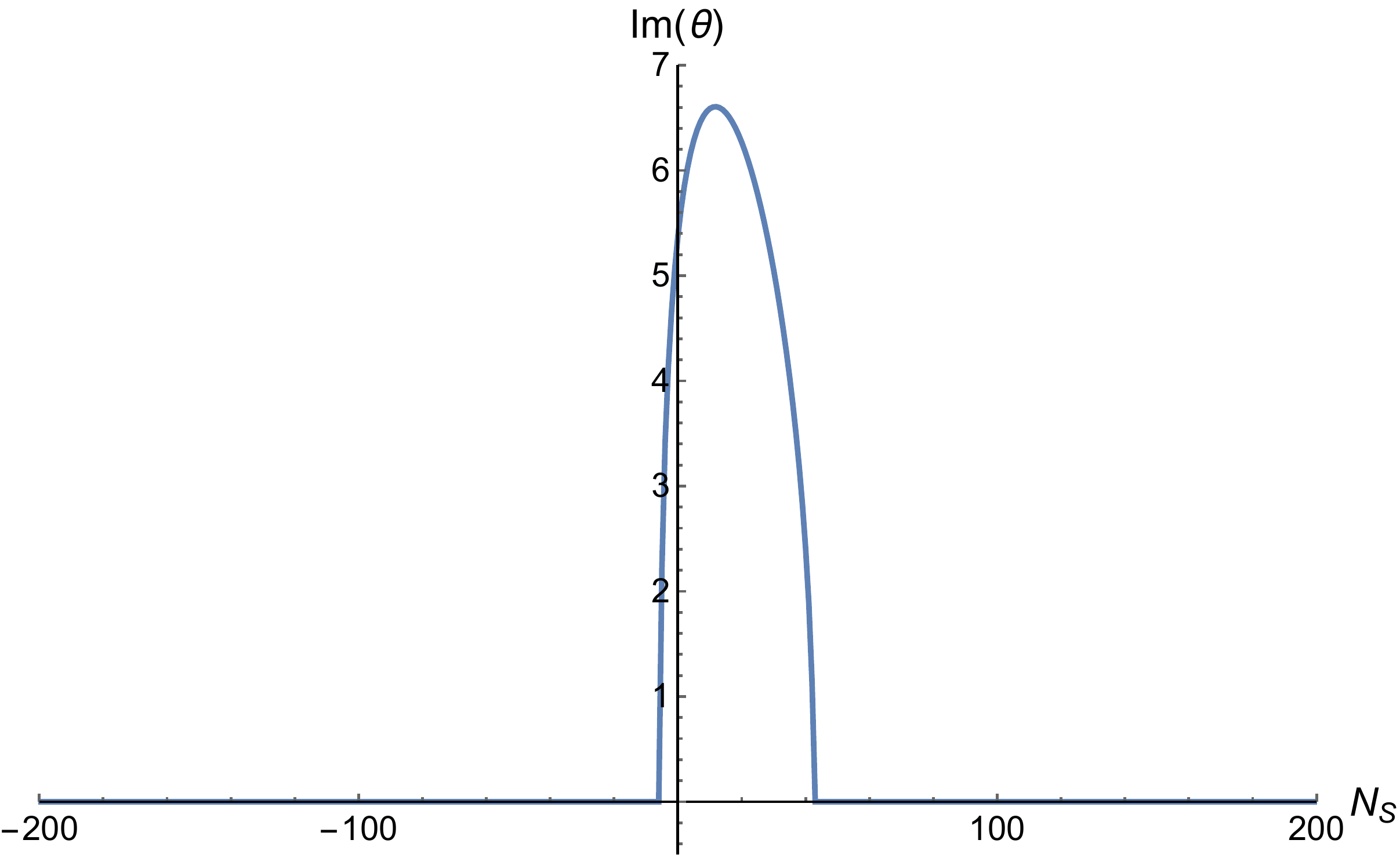} 
		\caption{Position (top) and stability coefficients (bottom) of the UV NGFPs appearing in gravity-scalar systems as a function of $N_S$. The fixed point structure undergoes qualitative changes at $N_S \approx -6$ and $N_S \approx 46$ where the critical exponents change from real to complex values.\label{scalarfp}} 
	\end{center}	 
\end{figure}
Notably, all values $N_S$ give rise to a NGFP with two UV attractive eigendirections. The position $(\lambda_*, g_*)$ and stability coefficients of this family of fixed points is displayed in Fig.\ \ref{scalarfp}. The first noticeable feature is a sharp transition in the position of the NGFP occurring at $N_S \simeq -6$: for $N_S \le -6$ the NGFP is located at $\lambda_* < 0$ while for $N_S > -5$ one has $\lambda_* > 0$.  For $N_S \rightarrow \infty$ the fixed point approaches $C\equiv(1/2,0)$, which can be shown to be a fixed point of the beta functions \eqref{betafunction} in the large $N_S$ limit. The value of the critical exponents shown in the lower line of Fig.\ \ref{scalarfp} indicates that there are two transitions: for $N_S \le -6$ there is a UV-attractive NGFP with two real critical exponents $\theta_1 \simeq 4$ and $\theta_2 \simeq 2$. These values are essentially independent on $N_S$. On the interval $-6 \le N_S \le 46$ the critical exponents turn into a complex pair. In particular for $N_S = 0$, one recovers the pure gravity fixed point NGFP \eqref{NGFP1}. For $N_S > 46$ one again has a UV-attractive NGFP with two real critical exponents with one of the critical exponents becoming large. Thus we clearly see a qualitatively different behavior of the NGFPs situated in the upper-right quadrant (relevant for GUT-type matter models) and the NGFPs in the lower-left quadrant (relevant for the standard model) of Fig.\ \ref{Fig.7}, reconfirming that the Asymptotic Safety mechanism realized within these classes of models is of a different nature.

\section{Summary and outlook}
\label{sect.7}

This work uses the functional renormalization group equation (FRGE) for the effective average action $\Gamma_k$ \cite{Wetterich:1992yh,Morris:1993qb,Reuter:1993kw,Reuter:1996cp} adapted to the Arnowitt-Deser-Misner (ADM) formalism \cite{Manrique:2011jc,Rechenberger:2012dt} to study the renormalization group flow of Newton's constant and the cosmological constant for minimally coupled gravity-matter models. As an important conceptual advantage the resulting construction equips spacetime with a natural foliation structure.  
The resulting distinguished ``time''-direction may be used to implement a Wick rotation from Euclidean to Minkowski signature.
  
The ADM-formalism expresses the metric degree's of freedom in terms of a Lapse function, a shift vector and a metric measuring distances on spatial slices, see eq.\ \eqref{fol1}. The key difficulty in obtaining a well-defined off-shell flow equation for this case originates from the fact that the Lapse function and the shift vector appear as Lagrange multipliers. Implementing ``proper-time gauge'' \cite{Dasgupta:2001ue}, using the freedom of choosing a coordinate system to eliminating the fluctuations in the Lapse function and shift vector,  leads to non-canonical propagators for the remaining fluctuation fields, see Table \ref{Tab.1}. Following \cite{Biemans:2016rvp}, our work bypasses this obstruction by implementing a new Feynman-type gauge fixing for the ADM-fields. 
The main virtue of the construction is that all fields, including the Lagrange multipliers and ghosts, obtain regular, relativistic dispersion relations. Moreover, all component fields propagate with the same speed of light when the dispersion relations are evaluated in a Minkowski background. This condition  fixes the gauge choice uniquely up to a physically irrelevant ${\mathbb Z}_2 \times {\mathbb Z}_2$ symmetry. The construction is reminiscent to Feynman gauge in quantum electrodynamics where the gauge-fixing provides a suitable kinetic term for the time-component of the gauge-potential.

In this work we apply the resulting flow equation to study the scale-dependence of Newton's constant and the cosmological constant for gravity minimally coupled to an arbitrary number of free scalar, vector, and Dirac fields. In this case, it suffices to evaluate the general FRGE on a flat Friedmann-Robertson-Walker background. 
The beta functions encoding the projected flow are encoded in the volume factor and extrinsic curvature terms constructed from the background. In this way our construction bypasses on of the main limitations of the Matsubara-type computations \cite{Manrique:2011jc,Rechenberger:2012dt} where time-direction was taken compact.

Our central result for the case of pure gravity is the phase diagram shown in Fig.\ \ref{flow2d3d}. Structurally, the result matches the phase diagrams obtained from studying similar RG flows in the metric formulation \cite{Reuter:2001ag,Litim:2003vp,Donkin:2012ud,Nagy:2013hka} and
from the evaluation of Lorentzian RG flows based on the Matsubara-formalism \cite{Manrique:2011jc,Rechenberger:2012dt}. In particular, 
we recover the key element of Asymptotic Safety, a UV-attractive non-Gaussian fixed point (NGFP). 
Subsequently, we classify the fixed point structure for gravity minimally coupled to an arbitrary number of free matter fields. We observe that the contribution of the matter sector can be encoded in a two-parameter deformation of the beta functions resulting from the pure gravity case and we give an explicit map between the field content of the model and the deformation parameters. In terms of the deformation parameters, it is found that the occurrence of a NGFP suitable for Asymptotic Safety is rather generic (see Fig. \ref{Fig.7}). In particular the field content of the standard model (and also its most commonly studied extensions) gives rise to a UV fixed point with real critical exponents. Moreover, our classification reveals that certain models with a low number of massless matter fields also admit an additional infrared fixed point which could provide the completion of the RG flow at low energy. 
 Our findings complement earlier studies based on the metric formalism \cite{Dona:2013qba,Meibohm:2015twa} by clearly demonstrating that the NGFPs responsible for Asymptotic Safety appearing for gravity coupled to the matter content of the standard model and grand unified type theories are qualitatively different. 

The setup developed in this work provides an important stepping stone for future developments of the gravitational Asymptotic Safety program. Conceptually, the results reported in this work remove one of the main obstructions for computing transition amplitudes between spatial geometries at different instances in time. Moreover, they provide a solid starting ground for computing real time correlation functions by combining the FRGE for the ADM formalism with the ideas advocated in \cite{Floerchinger:2011sc,Pawlowski:2015mia}. On the phenomenological side, our work evaluates the flow equation \eqref{FRGE} on a flat Friedmann-Robertson-Walker background and captures the gravitational fluctuations through the component fields typically used in cosmic perturbation theory. This setup is easily extended by including scalar fields and we showed explicitly that the Asymptotic Safety mechanism remains intact for this case. These features make the present framework predestined for studying the scale-dependence of cosmic perturbations within the Asymptotic Safety program also in the context of single-field inflationary models. We hope to come back to these points in the future. \\[2ex]

\paragraph*{Acknowledgements.}
We thank N.\ Alkofer, A.\ Bonanno, W.\ Houthoff, A.\ Kurov, R.\ Percacci,  M.\ Reuter, and C.\ Wetterich for helpful discussions. The research of F.~S.
is supported by the Netherlands Organisation for Scientific
Research (NWO) within the Foundation for Fundamental Research on Matter (FOM) grants 13PR3137 and 13VP12.
\begin{appendix}
\section{The flat Friedmann-Robertson-Walker background}
\label{App.A}
Throughout this work, we evaluate the flow equation on a flat (Euclidean)  Friedmann-Robertson-Walker background 
\be\label{FRWback2}
\gb_{\mu\nu} = {\rm diag} \left[ \, 1 \, , \, a(\tau)^2 \, \delta_{ij}\right] \qquad \Longleftrightarrow \qquad 
\Nb = 1 \, , \quad \Nb_i = 0 \, , \quad  \sib_{ij} = a(\tau)^2 \,  \delta_{ij} \, .  
\ee
In this background the projectors \eqref{proj1} take a particularly simple form
 \be\label{proj2}
 t^\mu = \big(   \, 1 \, , \, \vec{0} \, \big) \, , \qquad
 e_i{}^\mu = \big(  \vec{0} \, , \, \delta_i^j \big) \, , 
 \ee
 implying that $t^\mu$ is always normal to the spatial hypersurface $\Sigma_\tau$.
%
%
%
The extrinsic and intrinsic curvature tensors of this background satisfy
\be\label{curvatures}
\Kb_{ij} = \tfrac{1}{d} \, \Kb \, \sib_{ij} \, , \qquad \Rb = 0 \, , \qquad \Db_i = \p_i
\ee
where $\Kb \equiv \sib^{ij} \, \Kb_{ij}$. Moreover, the Christoffel-connection on the spatial slices vanishes such that $\Db_i = \p_i$.

In order to evaluate the operator traces appearing in the flow equation it is useful to resort to heat-kernel techniques with respect to the background spacetime \eqref{FRWback2}. For this purpose, we observe that \eqref{proj2} entails that there is a canonical ``lifting'' of vectors tangent to the spatial slice to $D$-dimensional vectors
\be\label{uplift}
v^i(\tau,y) \quad \mapsto \quad v^\mu(\tau,y) \equiv (0 \, , \, v^i(\tau,y))^{\rm T} \, . 
\ee
The $D$-dimensional Laplacian $\Delta_s \equiv - \gb^{\mu\nu} \Db_\mu \Db_\nu$ ($s=0,1,2$) naturally acts on these $D$-vectors. In order to rewrite the variations in terms of $D$-covariant quantities,  we exploit that $\Delta_s$ can be expressed in terms of the flat space Laplacian $\square \equiv - \partial_\tau^2 - \sib^{ij} \partial_i \partial_j$ and the extrinsic curvature.
 For the Laplacian acting on $D$-dimensional fields with zero, one, and two  indices one has
\be\label{LapDala}
\begin{split}
	\Delta_0 \phi = & \, \Big( \square - \Kb \partial_\tau \Big) \phi \, , \\
	 \Delta_1 \phi_\mu  = & \, \Big( \square - \tfrac{d-2}{d} \, \Kb \partial_\tau + \tfrac{1}{d} (\partial_\tau \Kb) + \tfrac{1}{d} \Kb^2 \Big) \phi_\mu  \, , \\
	\Delta_2 \phi_{\mu\nu} = & \,\Big( \square - \tfrac{d-4}{d} \, \Kb \partial_\tau + \tfrac{2}{d} (\partial_\tau \Kb) + \tfrac{2(d-1)}{d^2} \Kb^2 \Big)  \phi_{\mu\nu}  \, . \\
\end{split}
\ee
%
 When evaluating the traces by covariant heat-kernel methods, we then use the embedding map \eqref{uplift} together with the completion \eqref{LapDala} to express the operator $\square$ in terms of $\Delta_i$.

The operator traces appearing in \eqref{FRGE} are conveniently evaluated using standard heat-kernel formulas for the $D$-dimensional Laplacians \eqref{LapDala}
\be\label{heat1}
{\rm Tr}_i  \, e^{-s \left( \Delta_i + E \right) } \simeq \frac{1}{(4\pi s)^{D/2}} \int d^Dx \sqrt{g} \,  \Big[ {\rm tr}_i \, \unit + s \left( \tfrac{1}{6} \, {}^{(D)}R \, {\rm tr}_i \, \unit - {\rm tr}_i \, E \right) + \ldots \Big] \, . 
\ee
Here ${\rm tr}_i$ is a trace over the internal space and the dots indicate terms build from four and more covariant derivatives, which do not contribute to the present computation. For the FRW background, the spacetime curvature ${}^{(D)}R$ can readily be replaced by the extrinsic curvature, evoking
\be\label{rep1}
\int d^Dx \sqrt{\gb} \; {}^{(D)}R = \int d\tau d^dy \sqrt{\sib} \left[ \tfrac{d-1}{d} \Kb^2 \right] \, . 
\ee 

Combining the diagonal form of the projectors \eqref{proj2} with the $D$-dimensional heat-kernel expansion \eqref{heat1} allows to write operator traces for the component fields. On the flat FRW background these have the structure
\be\label{heat2}
{\rm Tr}_i \, e^{-s \Delta_i} = \frac{1}{(4\pi s)^{D/2}} \, \int d\tau d^dy \sqrt{\sib} \, \Big[  a_0  + a_2 \, s \, \Kb^2 + \ldots  \Big] \, , 
\ee
The coefficient $a_n$ depend on the index structure $i$ of the fluctuation field and are listed in Table \ref{Tab.heat}. 
\begin{table}[t!]
	\renewcommand{\arraystretch}{1.4}
	\begin{center}
\begin{tabular}{|c|c|c|c|c|c|}
\hline \hline
      & $S$ & $V$ & $T$ & $TV$ & $TTT$ \\ \hline
$a_0$ & $1$ & $d$ & $\tfrac{1}{2}d(d+1)$     & $d-1$ & $\tfrac{1}{2}(d+1)(d-2)$ \\
$a_2$ & $\frac{d-1}{6d}$ & $\frac{d-1}{6}$ & $\frac{(d-1)(d+1)}{12}$ & 
$\frac{d^3-2d^2+d+6}{6d^2}$ & $\frac{d^4-2d^3-d^2+14d+36}{12d^2}$ \\
\hline \hline
\end{tabular}
\caption{\label{Tab.heat} Heat-kernel coefficients for the component fields appearing in the decompositions \eqref{TTshift} and \eqref{TTmet}. Here $S$, $V$, $T$, $TV$, and $TTT$ are scalars, vectors, symmetric two-tensors, transverse vectors, and transverse traceless symmetric matrices, respectively. }
\end{center}
\end{table}
This result \eqref{heat2} is the key ingredient for evaluating the operator traces of the flow equation on a flat FRW background. 

\section{Hessians in a Friedmann-Robertson-Walker background}
\label{App.B}
The evaluation of the flow equation \eqref{FRGE} requires the Hessian $\Gamma_k^{(2)}$. The technical details of this calculation
are summarized in this appendix. In the sequel, indices are raised and lowered with the background metric $\sib_{ij}$. Moreover, we introduce the shorthand notations
\be\label{shorthand}
\int_x \equiv \int \, d\tau \, d^dy \, \sqrt{\sib} \; , \qquad \mbox{and} \qquad \sh \equiv \sib^{ij} \, \sh_{ij}
\ee
to lighten the notation and use $\Delta \equiv - \sib^{ij} \p_i \p_j$ to denote the Laplacian on the spatial slices. 

\subsection{Hessians in the gravitational sector: decomposition of fluctuations}
\label{App.B1}
When constructing $\Gamma_k^{(2)}$, it is convenient to consider \eqref{GammaEH} as a linear combination of the interaction monomials \eqref{imon}. These monomials are then expanded in terms of the fluctuation fields according to
\be
N = \Nb + \Nh \, , \qquad N_i = \Nb_i + \Nh_i \, , \qquad \sigma_{ij} = \sib_{ij} + \sh_{ij} \, .
\ee
As an intermediate result, we note that the expansion of the extrinsic curvature \eqref{Kext} around the FRW background is given by
\be\label{extvar}
\begin{split}
\delta K_{ij} = & \, -  \hat{N} \, \bar{K}_{ij} + \tfrac{1}{2} ( \partial_{\tau} \sh_{ij} -\p_i \hat{N}_j - \p_j \hat{N}_i )	\, , \\
\delta^2 K_{ij} = & \, 
{2}\, \hat{N}^2 \, \bar{K}_{ij} 
- \hat{N} \left( \partial_{\tau} \sh_{ij} -\p_i \hat{N}_j - \p_j \hat{N}_i \right)
+ \hat{N}^k \left( \partial_i \sh_{jk} + \p_j \sh_{ik} - \p_k \sh_{ij} \right) \, , 
\end{split}
\ee
were $\delta^n$ denotes the order of the expression in the fluctuation fields. 
For later reference, it is also useful to have the explicit form of these expressions contracted withthe inverse background metric
\be\label{traceextvar}
\begin{split}
	\sib^{ij} \left( \delta K_{ij} \right) = & \, - \Nh \Kb + \half \sib^{ij} (\p_\tau \sh_{ij}) - \p^i \Nh_i \, , \\
	\sib^{ij} \left(\delta^2 K_{ij} \right) = &  \, {2} \, \Nh^2 \Kb - \Nh \sib^{ij}  \left( \p_\tau \sh_{ij} \right) + 2 \Nh \p^i \Nh_i + \Nh^k \left(2 \, \p^i \, \sh_{ik} - \p_k \,  \sh \right) \, . 
\end{split}
\ee

Expanding the interaction monomials \eqref{imon}, the terms quadratic in
the fluctuation fields are 
\be\label{varI1}
\begin{split}
	\delta^2 I_1 = & \int_x \Big[ 
	2 (\delta K_{ij}) \sib^{ik} \sib^{jl} (\delta K_{kl}) 
	+ \tfrac{2}{d} \, \Kb \, \sib^{ij} \left( \delta^2 K_{ij} + (\delta K_{ij}) (2 \Nh + \sh ) \right) \\ & \qquad
	- \tfrac{8}{d} \, \Kb \, \sh^{ij} \, (\delta K_{ij}) 
	+ \tfrac{1}{d} \, \Kb^2 \left( \tfrac{d-4}{d} \Nh \sh + \tfrac{d-8}{4d} \sh^2 - \tfrac{d-12}{2d} \sh_{ij} \sh^{ij} \right)
	 \Big] \, , \\ 
	\delta^2 I_2 = & \int_x \Big[ 2 (\sib^{ij} \, \delta K_{ij})^2 + 2  \Kb \, \sib^{ij} \left( \delta^2 K_{ij} + (\delta K_{ij}) (2 \Nh + \sh ) \right) - 4 \Kb \sh^{ij} \, (\delta K_{ij}) \\ & \qquad
	+ \Kb^2 \left( \tfrac{d-4}{d} \Nh \sh + \tfrac{d^2-8d+8}{4d^2} \sh^2 - \tfrac{d-8}{2d}  \sh_{ij} \sh^{ij}  \right) 
	{ - \tfrac{4}{d} \Kb \sh \sib^{ij} \delta K_{ij}}  \Big] \, , \\
	\delta^2 I_3 = & \, \int_x \left[ \big( 2 \Nh + \sh \big)\big(\p_i \p_j \sh^{ij} + \Delta \sh \big) - \tfrac{1}{2} \sh_{ij} \Delta \sh^{ij} - \tfrac{1}{2} \sh \Delta \sh + \big(\p_i \sh^{ik}\big) \big(\p_j \sh^j{}_k \big) \right] \, , \\
	\delta^2 I_4 = & \, \int_x \left[ \Nh \sh + \tfrac{1}{4} \sh^2 - \tfrac{1}{2} \sh^{ij} \sh_{ij} \right] \, .
\end{split}
\ee
In order to arrive at the final form of these expressions, we integrated by parts and made manifest use of the geometric properties of the background \eqref{curvatures}. 

In order to develop a consistent gauge-fixing scheme and to simplify the structure of the flow equation it is useful to carry out a further transverse-traceless decomposition of the fluctuation fields entering into
\eqref{varI1}. A very convenient choice is provided by the standard decomposition of the fluctuation fields used in cosmic perturbation theory (see, e.g., \cite{Baumann:2009ds} for a detailed discussion) where the shift vector and the metric on the spatial slice are rewritten according to
\be\label{TTdec}
\begin{split}
\Nh_i = & \, u_i + \p_i \, \tfrac{1}{\sqrt{\Delta}} \, B \, , \\
\sh_{ij} = & \, h_{ij} - \left( \sib_{ij} + \p_i \p_j \, \tfrac{1}{\Delta} \right)  \psi + \p_i \p_j \, \tfrac{1}{\Delta} \, E + \p_i \tfrac{1}{\sqrt{\Delta}} v_j + \p_j \, \tfrac{1}{\sqrt{\Delta}} \, v_i \, .
\end{split}
\ee
The component fields are subject to the constraints
\be
\p^i \, u_i = 0 \, , \qquad \p^i \, h_{ij} = 0 \, , \quad \sib^{ij} h_{ij} = 0 \, , \quad \p^i v_i = 0 \, , 
\ee
indicating that $u_i$ and $v_i$ are transverse vectors and $h_{ij}$ is a transverse-traceless tensor. Notably, the partial derivatives and $\Delta$ can be commuted freely, since the background metric is independent of the spatial coordinates. The normalization of the component fields has been chosen such that the change of integration variables does not give rise to non-trivial Jacobians. This can be seen from noting
\be\label{aux1}
\begin{split}
	\Nh_i \, \Nh^i = & \,  u_i \, u^i + B^2 \, , \\
	\sh_{ij} \, \sh^{ij} = & \, h_{ij} h^{ij} + (d-1) \, \psi^2 + E^2 + 2 \, v_i \, v^i \, , \\
\end{split}
\ee
implying that a Gaussian integral over the ADM fluctuations leads to a Gaussian integral in the component fields which does not give rise to operator-valued determinants.

The final step expresses the variations \eqref{varI1} in terms of the component fields \eqref{TTdec}. The rather lengthy computation can be simplified by using the identities
\be\label{aux2}
\begin{split}
	\sh =  - (d-1) \psi - E \, , \qquad
	\p^i \sh_{ij} =  - \p_j E - \sqrt{\Delta} \, v_j \, , \qquad
	\p^i \p^j \sh_{ij} = \Delta E \, . 
\end{split}
\ee
together with the relations \eqref{aux1} and \eqref{aux2}. 
Starting with the kinetic terms appearing in 
 $\delta^2 I_1$ and $\delta^2 I_2$,
\be\label{kinterms}
{\Kbb}_1 \equiv \, 2 \int_x \,  
(\delta K_{ij}) \, \sib^{ik} \sib^{jl} \, (\delta K_{kl}) \, , \qquad 
{\Kbb}_2 \equiv  \, 2 \int_x (\sib^{ij} \, \delta K_{ij})^2 \, ,
\ee
the resulting expressions written in terms of component fields are
\be
\begin{split}
	\Kbb_1 = & \, \int_x \Big[ 
	- \half \, h^{ij} \left( \p_\tau^2 + \tfrac{d-4}{d} \Kb \p_\tau \right) h_{ij}
		- \tfrac{d-1}{2} \, \psi \left(\p_\tau + \tfrac{d-2}{d} \Kb \right) \left( \p_\tau + \tfrac{2}{d} \Kb \right) \psi \\ & \qquad \; 		
	- \half E \left(\p_\tau + \tfrac{d-2}{d} \Kb \right) \left( \p_\tau + \tfrac{2}{d} \Kb \right) E 
	- v^i \left( \p_\tau + \tfrac{d-3}{d} \Kb \right) \left( \p_\tau + \tfrac{1}{d} \Kb \right) v_i \\ & \qquad \;
	- 2 \, B \, \sqrt{\Delta} \, \left( \p_\tau + \tfrac{2}{d} \Kb \right) E
	- 2 \, u^i \left( \p_\tau + \tfrac{2}{d} \Kb \right) \sqrt{\Delta} \, v_i + 2 B \Delta B \\ & \qquad \;
	+ u_i \Delta u^i - \tfrac{4}{d} \Kb \Nh \, \sqrt{\Delta} \, B
	+ \tfrac{2}{d} \Kb \Nh \, \left( \p_\tau + \tfrac{2}{d} \Kb \right) \, \big((d-1) \psi + E \big) \\ & \qquad \;
	+ \tfrac{2}{d} \, \Kb^2 \, \Nh^2 
	\Big] \, , 
\end{split}
\ee
and
\be
\begin{split}
	\Kbb_2 = & \, \int_x \Big[ - \tfrac{1}{2} \big((d-1)\psi + E\big) \left(\p_\tau + \tfrac{d-2}{d} \Kb \right) \left( \p_\tau + \tfrac{2}{d} \Kb \right) \big((d-1)\psi + E\big) \\ & \qquad \; 
	-2 B \sqrt{\Delta} \left( \p_\tau + \tfrac{2}{d} \Kb \right) \big((d-1)\psi + E\big) + 2 B \Delta B  \\
	& \qquad \; 
	  - 4 \Kb \Nh \sqrt{\Delta} \, B  
	+ 2 \Kb \Nh \left( \p_\tau + \tfrac{2}{d} \Kb \right) \big((d-1)\psi + E\big)
	+ 2 \Kb^2 \Nh^2
	\Big] \, . 
\end{split}
\ee
On this basis one finds that
\be\label{I1res}
\begin{split}
	\delta^2 I_1 =  \Kbb_1  - \, \int_x \Big[ \; & + \tfrac{(d-4)(d-8)}{4d^2} \, \Kb^2 \Nh \, \big( (d-1)\psi + E \big) \\
	&   	
	 + \tfrac{4}{d} \, \Kb \, h^{ij} \big( \p_\tau + \tfrac{d-12}{8d} \Kb \big) h_{ij} \\
	& 
	- \tfrac{4}{d} \, \Kb \, \big( u^k \, \sqrt{\Delta} \, v_k + B \, \sqrt{\Delta} \, E - v^i \, ( \p_\tau + \tfrac{d-8}{4d} \, \Kb) \, v_i \big) \\
	&  
	- \tfrac{1}{d} \, \Kb \, \big((d-1) \psi + E \big) \big(\p_\tau + \tfrac{1}{4} \Kb \big) \big((d-1) \psi + E \big) \\
	&  
	+ \tfrac{4}{d} \, \Kb \, E \, \big(\p_\tau + \tfrac{d+4}{8d} \, \Kb \big) \, E
	+ \tfrac{4(d-1)}{d} \, \Kb  \, \psi  \,\big(\p_\tau + \tfrac{d+4}{8d} \, \Kb \big) \, \psi
	 \Big]
\end{split}
\ee
and
\be\label{I2res}
\begin{split}
	\delta^2 I_2 =  \Kbb_2  - \, \int_x \Big[ \; & - \tfrac{d-4}{d} \, \Kb^2 \Nh \big( (d-1) \psi + E \big) + 2 \, \Kb \, h^{ij} \, \big(\p_\tau +\tfrac{d-8}{4d} \, \Kb \big) \, h_{ij} \\
	& 		
     - \Kb \, \big((d-1) \psi + E \big) \big(\tfrac{d-2}{d} \, \p_\tau + \tfrac{d^2-8}{4d^2} \,     \Kb \, \big) \big((d-1) \psi + E \big) \\
	&  		
	+ 2\,  \Kb \, E \, \big(\p_\tau +\tfrac{1}{4} \, \Kb \big) \, E 
	+ 2 (d-1) \Kb \, \psi \, \big(\p_\tau +\tfrac{1}{4} \, \Kb \big) \, \psi \\
	&  		
	+2 \Kb v^i \, \big(\p_\tau +\tfrac{d-6}{2d} \, \Kb \big) \, v_i  {- \tfrac{4}{d} \Kb \left( (d-1) \psi + E \right)  \sqrt{\Delta} B }
	 \Big] \, . 
\end{split}
\ee 
Finally, $\delta^2 I_3$ and $\delta^2 I_4$ written in terms of the component fields are
\be\label{I3res}
\begin{split}
	\delta^2 I_3 = & \, \int_x \Big[ 
	\tfrac{ (d-1) (d-2)}{2} \, \psi \Delta \psi - \half \, h_{ij} \Delta h^{ij} - 2 \, (d-1) \, \Nh \Delta \psi 
	\Big]  , \\
	\delta^2 I_4 = &  \int_x \Big[ \tfrac{(d-1)(d-3)}{4} \psi^2 + \tfrac{(d-1)}{2} \psi  E - \tfrac{1}{4} E^2 
	 - \Nh \big( (d-1) \psi + E  \big)  - \half  h_{ij} h^{ij} - v_i v^i  \Big]  . 
\end{split}
\ee
Combining these variations according to \eqref{GammaEH} one 
arrives the matrix entries for $\delta^2\Gamma_k^{\text{grav}}$. In the flat space limit, where $\bar{K}=0$, these entries are listed in the second column of Table \ref{Tab.1}.
\subsection{Gauge-fixing terms}
\label{App.B3}
Following the strategy of the previous subsection it is useful to also decompose the gauge-fixing terms \eqref{gf:ansatz} into two interaction monomials
\be\label{I56def}
\delta^2 I_5 \equiv \int_x \, F^2 \, , \qquad \delta^2 I_6 \equiv \int_x F_i \, \sib^{ij} \, F_j \, . 
\ee
Since the functionals $F$ and $F_i$ defined in eq.\ \eqref{gauge1} are linear in the fluctuation fields, the gauge-fixing terms are quadratic in the fluctuations by construction. This feature is highlighted by adding the $\delta^2$ to the definition of the monomials. 

Substituting the explicit form of $F$ and $F_i$ and recasting the resulting expressions in terms of the component fields \eqref{TTdec}
one finds
\be\label{d2I5}
\begin{split}
\delta^2 I_5 = & \, \int_x \Big[ c_2^2 \, B \Delta B
- \Nh \, \left(c_1 \p_\tau + (c_1 - c_9) \Kb \right) \left( c_1 \p_\tau + c_9 \Kb \right) \Nh \\ & \quad
-  \big((d-1)\psi + E\big) \left(c_3 \p_\tau + (c_3 - c_8) \Kb \right) \left( c_3 \p_\tau + c_8 \Kb \right) \big((d-1)\psi + E\big) \\ & \quad
- 2 \, c_2 \, B \, \sqrt{\Delta} \, \left(c_1 \, \p_\tau + c_9 \Kb \right) \, \Nh 
\\ & \quad
+ 2 \Nh \, \left(c_1 \p_\tau + (c_1-c_9) \Kb \right) \left(c_3 \p_\tau + c_8 \Kb \right) \big((d-1)\psi + E\big)
\\ & \quad
+2 \, c_2 \, B \sqrt{\Delta} \, \left(c_3 \p_\tau + c_8 \Kb \right) \, \big((d-1)\psi + E\big)
 \Big]	
\end{split}
\ee
and
\be\label{d2I6}
\begin{split}
	\delta^2 I_6 = & \, \int_x \Big[ c_5^2 \, \Nh \Delta \Nh 
	- \,  u^i \, \big(c_4 \, \p_\tau + ( \tfrac{d-2}{d} c_4 - c_{10} ) \, \Kb \big) \big( c_4 \p_\tau + c_{10} \Kb \big) \, u_i \\  & \; \; \qquad
	         - B \big(c_4 \, \p_\tau + ( \tfrac{d-1}{d} c_4 - c_{10} ) \, \Kb \big) \big(  c_4 \p_\tau + (\tfrac{1}{d} \, c_4 + c_{10}) \, \Kb \big) \,  B  
	         \\  & \; \; \qquad 
	+2 \, c_5 \, \Nh \big( c_4 \p_\tau + (\tfrac{2}{d} c_4 + c_{10}) \,  \Kb \big) \sqrt{\Delta} B \\  & \; \; \qquad
-2 \, c_6 \, \big((d-1)\psi + E\big) \big(c_4 \p_\tau + (\tfrac{2}{d} c_4 + c_{10}) \Kb \big)  \sqrt{\Delta} B
	\\  & \; \; \qquad
		-2 \, c_7 \,  E \big(c_4 \p_\tau + (\tfrac{2}{d} c_4 +c_{10}) \Kb \big) \sqrt{\Delta} B \\  & \; \; \qquad
		- 2 \, c_7 \, v^i \sqrt{\Delta} \, ( c_4 \p_\tau + c_{10} \Kb ) \, u_i 
		-2 c_5 c_6 \, \Nh \Delta \, \big((d-1)\psi + E\big)
	\\  & \; \; \qquad     
	-2 \, c_5 \, c_7 \, \Nh \Delta E 
	+ c_6^2 \, \big((d-1)\psi + E\big) \Delta \big((d-1)\psi + E\big)
	\\  & \; \; \qquad
	+ 2 c_6 c_7 \, \big((d-1)\psi + E\big) \Delta E 
	+ c_7^2 \, \big(E\Delta E + v^i \Delta v_i \big)    
	\Big]	\, . 
\end{split}
\ee
Here again we made use of the geometric properties of the background and integrated by parts in order to obtain a similar structure as in the gravitational sector.

Combining the results \eqref{I1res}, \eqref{I2res}, \eqref{I3res}, \eqref{d2I5}, and \eqref{d2I6}, taking into account the relative signs between the terms and restoring the coupling constants according to \eqref{GammaEH2} gives the part of the gauge-fixed gravitational action quadratic in the fluctuation fields. The explicit result is rather lengthy and given by 
\be\label{eqS1B}
\begin{split}
	& \,	32  \pi G_k  \Big( \half \delta^2 \Gamma^{\rm grav}_k + \Gamma_k^{\rm gf} \Big) = \\ & \, 
	\int_x \Big\{ - \Nh \left[ (c_1 \p_\tau +(c_1 - c_9) \Kb)(c_1 \p_\tau + c_9 \Kb) - c_5^2 \, \Delta + {\tfrac{2(d-1)}{d} \, \Kb^2} \right] \Nh \\ 	&  \qquad 
	- B \, \left[  (c_4 \p_\tau + ( \tfrac{d-1}{d} c_4 - c_{10}) \Kb)(c_4  \p_\tau +( \tfrac{1}{d} c_4 + c_{10})\Kb) - c_2^2 \, \Delta \right] B \\ &  \qquad
	- 2 \, B \, \sqrt{\Delta} \left[ (c_1 c_2 + c_4 c_5) \p_\tau
	+ (c_2 c_9 + c_4 c_5 \, \tfrac{d-2}{d} - c_5 c_{10} - 
	\tfrac{2(d-1)}{d}) \Kb \right] \Nh \\ &  \qquad
	+ 2 \, \Nh \Big[
	(c_1 \p_\tau + (c_1 - c_9)\Kb)(c_3 \p_\tau + c_8 \Kb) - \tfrac{d-1}{d} \Kb \p_\tau
	- c_5 (c_6 + c_7) \Delta \\ &  \qquad \qquad \quad
	- {\tfrac{5d^2-12d+16}{8d^2}} \Kb^2 -  \Lambda_k 
	\Big] E \\ &  \qquad
	+ 2 (d-1) \, \Nh \Big[
	(c_1 \p_\tau + (c_1 - c_9) \Kb)(c_3 \p_\tau +c_8 \Kb) - \tfrac{d-1}{d} \Kb \p_\tau \\ &  \qquad \qquad \quad
	+ (1 - c_5 c_6) \Delta  - {\tfrac{5d^2-12d+16}{8d^2}} \Kb^2 -  \Lambda_k 
	\Big] \psi \\ &  \qquad
	+ 2 \, B \sqrt{\Delta} \left[ \big(c_2 c_3 + c_4(c_6+c_7) \big) \p_\tau +
	\big( c_2 c_8 + (c_6 + c_7)(\tfrac{d-2}{d} c_4 - c_{10}) \big) \Kb \right] E \\ &  \qquad
	+2 (d-1) B \sqrt{\Delta} \left[
	\big(1 + c_2c_3+c_4c_6\big) \p_\tau + \big( c_2 c_8 + \tfrac{d-2}{d} c_4 c_6 - c_6 c_{10} \big) 
	\Kb\right] \psi \\ &  \qquad
	- (d-1) \, \psi \, \Big[	 (d-1)\left((c_3 \p_\tau + (c_3 -c_8)\Kb)(c_3 \p_\tau + c_8 \Kb) - c_6^2 \Delta \right) 
	\\ & \qquad \qquad \quad
	+ \tfrac{d-2}{2} ( - \p_\tau^2 + \Delta - {\tfrac{2}{d} } \dot{\Kb} ) + 
	{\tfrac{d^2 - 10d +14}{2d} } \Kb \p_\tau + {\tfrac{d^2-8d+11}{4d} } \, \Kb^2  - \tfrac{d-3}{2} \Lambda_k
	\Big] \psi
	\\ &  \qquad
	+ E  \Big[ (c_6 +c_7)^2 \Delta -(c_3 \p_\tau + (c_3 - c_8) \Kb)(c_3 \p_\tau + c_8 \Kb)  
	\\ & \qquad \qquad \quad
	- \half \Lambda_k + { \tfrac{d-1}{d}} \Kb \p_\tau + {\tfrac{d-1}{4d}} \, \Kb^2 \Big]  E \\ &  \qquad
	+ (d-1) \, \psi \Big[	 2 c_6 (c_6 + c_7) \Delta -2 (c_3 \p_\tau + (c_3 -c_8)\Kb)(c_3 \p_\tau + c_8 \Kb)
	\\ & \qquad \qquad \quad
	+ \p_\tau^2  + {\Kb \p_\tau }  + {\tfrac{d-1}{d} \dot{\Kb} } + {\tfrac{d-1}{2d}} \Kb^2 + {\Lambda_k}
	\Big] E
	\\ &  \qquad
	- u^i \left[ \big(c_4 \p_\tau + (\tfrac{d-2}{d} c_4 -c_{10})\Kb\big) \big(c_4 \p_\tau + c_{10} \Kb \big) - \Delta\right] u_i \\ &  \qquad
	+ v^i \left[ - \p_\tau^2 + \tfrac{d-2}{d} \Kb \p_\tau - \tfrac{1}{d} \dot{\Kb} + \tfrac{d^2-8d+11}{d^2} \Kb^2 + c_7^2 \, \Delta - 2 \Lambda_k \right] v_i \\ &  \qquad
	- 2 \, u^i \left[ \big(1-c_4 c_7\big)   \p_\tau
	+ c_7 \big(  c_{10} - \tfrac{d-2}{d} c_4 \big) \Kb   \right] \, \sqrt{\Delta} \, v_i \\ &  \qquad
	+ \half \, h^{ij} \left[ -\p_\tau^2 + \tfrac{3d-4}{d} \, \Kb \p_\tau + \tfrac{d^2 - 9d+12}{d^2} \, \Kb^2 + \Delta - 2 \Lambda_k \right] \, h_{ij}
	\Big\} \, . 
\end{split} 
\ee

Based on this general result, one may then search for a particular gauge fixing which, firstly, eliminates all terms containing $\sqrt{\Delta}$ and, secondly, ensures that all component fields obey a relativistic dispersion relation in the limit when $\bar{K} = 0$. A careful inspection of eq.\ \eqref{eqS1B} shows that there is an essentially unique gauge choice which satisfies both conditions. The resulting values for the coefficients $c_i$ are given in eq.\ \eqref{gffinal}. Specifying the general result to these values finally results in the gauge-fixed Hessian appearing in the gravitational sector \eqref{eqS2Frank}. Taking the limit $\bar{K} = 0$, the propagators resulting from this gauge-fixing are displayed in the third column of Table \ref{Tab.1}. In this way it is straightforward to verify that the gauge choice indeed satisfies the condition of a relativistic dispersion relation for all component fields.

The gauge-fixing is naturally accompanied by a ghost action exponentiating the resulting Faddeev-Popov determinant. For the gauge-fixing conditions $F$ and $F_i$ the ghost action comprises a scalar ghost $\bar{c}, c$ and a (spatial) vector ghost $\bar{b}^i, b_i$. Their action can be constructed in a standard way by evaluating 
\be\label{ghs}
\Gamma_k^\text{scalar  ghost}=\int_x \bar{c} \, \, \frac{\delta F}{\delta \hat{\chi}^i} \, \, \delta_{c,b_i} \chi^i \; , \qquad 
\Gamma_k^\text{vector  ghost}=\int_x \bar{b}^{\, j} \, \, \frac{\delta F_j}{\delta \hat{\chi}^i} \, \, \delta_{c,b_i} \chi^i \, . 
\ee
Here $\frac{\delta F}{\delta \hat{\chi}^i}$ denotes the variation of the gauge-fixing condition with respect to the fluctuation fields $\hat{\chi} = \left[ \hat{N},\hat{N}_i,\hat{\sigma}_{ij} \right]$ at fixed background and
the expressions $\delta_{c,b_i} \chi^i$ are given by the variations \eqref{eq:gaugeVariations} with the parameters $f$ and $\zeta_i$ replaced by the scalar ghost $c$ and vector ghost $b_i$, respectively. Taking into account terms quadratic in the fluctuation fields only, the resulting ghost action is given in eq.\ \eqref{Gghost}. Together with the Hessian in the gravitational sector, eq.\ \eqref{eqS2Frank}, this result completes the construction of the Hessians entering the right-hand-side of the flow equation \eqref{FRGE}.

\section{Evaluation of the operator traces}
\label{App.C}
In Sect.\ \ref{sect.3} the operator traces have been written in terms of the standard $D = d+1$-dimensional Laplacian $\Delta_s \equiv - \gb^{\mu\nu} D_\mu D_\nu$ where $s = 0,1,2$ indicates that the Laplacian is acting on fields with zero, one or two spatial indices. In this appendix, we use the heat-kernel techniques  detailed, e.g., in \cite{Reuter:1996cp,Codello:2008vh} and \cite{Benedetti:2010nr} to construct the resulting contributions to the flow. 

\subsection{Cutoff scheme and master traces}
The final step in the construction of the right-hand-side of the flow equation is the specification of the regulator $\cR_k$. Throughout this work, we will resort to regulators of Type I, which are implicitly defined through the relation that the regulator dresses up each $D$-dimensional Laplacian by a scale-dependent mass term according to the rule
\be\label{Rscheme}
\Delta_s \mapsto P_k \equiv \Delta_s + R_k \, . 
\ee
Here  $R_k$ denotes a scalar profile function, providing the $k$-dependent mass term for the fluctuation modes. The prescription \eqref{Rscheme} then fixes the matrix-valued regulator $\cR_k$ uniquely. In this course, we first notice that the  matrix elements $\Gamma^{(2)}_k$ found in App.\ \ref{App.B} take the form
\be
\left. \Gamma^{(2)}_k \right|_{\hat{\chi}_i \hat{\chi}_j} = \left(32 \pi G_k \right)^{-\alpha_s} \, c \, \left[ \Delta_s + w + v_1 \, \Kb^2 + v_2 \, \dot{\Kb} + v_3 \, \Kb \p_\tau \right] \, , 
\ee
where $\alpha_s = 1,0$ depending on whether the matrix element arises from the gravitational or the ghost sector and $w$ encodes a possible contribution from a cosmological constant. Moreover, $c$ and the $v_i$ are $d$-dependent numerical coefficients whose values can be read off from eqs.\ \eqref{eqS2Frank} and \eqref{Gghost} Applying the rule \eqref{Rscheme} then yields
\be
\left. \cR_k \right|_{\hat{\chi}_i \hat{\chi}_j} = \left(32 \pi G_k \right)^{-\alpha_s} \, c \, R_k \, . 
\ee

Subsequently, one has to construct the inverse of $(\Gamma_k^{(2)} + \cR_k)$. Given the left-hand-side of the flow equation \eqref{FRGElhs} it thereby suffices to keep track of terms containing up to two time-derivatives of the background quantities, i.e., $\Kb^2$ and $\dot{\Kb}$. This motivates the split 
\be\label{PVdec}
 \left(\Gamma_k^{(2)} + \cR_k\right)  \equiv \cP + \cV \, , 
\ee 
where the propagator-matrix $\cP$ collects all terms containing $\Delta_s$ and $\Lambda_k$ and the potential-matrix $\cV$ collects the terms with at least one power of the extrinsic background curvature $\Kb$. The inverse $ (\Gamma_k^{(2)} + \cR_k)^{-1}$ can then be constructed as an expansion in $\cV$. Retaining terms containing up to two powers of $\Kb$ only 
\be\label{exp}
\big(\cP + \cV\big)^{-1} = \cP^{-1} - \cP^{-1} \, \cV \, \cP^{-1} +  \cP^{-1} \, \cV \, \cP^{-1} \, \cV \, \cP^{-1} + \cO(\Kb^3) \, . 
\ee

Typically, $\cP + \cV$ has a block-diagonal form in field space. At this stage it is instructive to look at a single block for which we assume that it is spanned by a single field (e.g., $h_{ij}$). In a slight abuse of notation we denote the propagator and potential on this block by $\cP$ and $\cV$ as well. From the structure of the Hessians one finds that the propagator has the form
\be\label{propagator}
\cP^{-1} = (32 \pi G_k)^{\alpha_s} \, c^{-1} \, \left( \Delta_s + R_k + w \right)^{-1}  \, ,
\ee 
while the potential $\cV$ is constructed from three different types of insertions
\be\label{pottypes}
\cV_1 = (32 \pi G_k)^{-\alpha_s} \, c \, \Kb^2 \, , \quad 
\cV_2 = (32 \pi G_k)^{-\alpha_s} \, c \, \dot{\Kb} \, , \quad
\cV_3 = (32 \pi G_k)^{-\alpha_s} \, c \, \Kb \p_\tau \, . 
\ee

The structure \eqref{exp} can be used to write the right-hand-side of the flow equation in terms of master traces, which are independent of the particular choice of cutoff function. Defining the profile function $R^{(0)}(\Delta_s/k^2)$ through the relation $R_k = k^2 R^{(0)}(\Delta_s/k^2)$, it is convenient to introduce the dimensionless threshold functions \cite{Reuter:1996cp}
\be
\begin{split}
	\Phi^p_n(w) \equiv \frac{1}{\Gamma(n)} \int_0^\infty dz \, z^{n-1} \, \frac{R^{(0)}(z) - z R^{(0)\prime}(z)}{[z+ R^{(0)}(z) + w]^p} \, , \\
	\widetilde{\Phi}^p_n(w) \equiv \frac{1}{\Gamma(n)} \int_0^\infty dz \, z^{n-1} \, \frac{R^{(0)}(z)}{[z+ R^{(0)}(z) + w]^p} \, .
\end{split}
\ee
For a cutoff of Litim type, $R_k=(k^2-\Delta_s)\,\theta(k^2-\Delta_s)$, to which we resort in the main part of the paper the integrals in the threshold functions can be evaluated analytically, yielding
\be\label{thresholdfcts}
\Phi^p_n(w) \equiv \frac{1}{\Gamma(n+1)} \, \frac{1}{(1+w)^p} \, , \qquad
\widetilde{\Phi}^p_n(w) \equiv \frac{1}{\Gamma(n+2)} \, \frac{1}{(1+w)^p} \, . 
\ee

The right-hand-side of the flow equation is then conveniently evaluated in terms of the following master traces. For zero potential insertions one has
\be\label{master0}
\begin{split}
{\rm Tr}\left[\cP^{-1} \p_t \cR_k \right] = \tfrac{k^{D}}{(4\pi)^{D/2}} \int_x & \Big[ a_0 \left( 2  \Phi^1_{D/2}(\tilde{w}) - \eta \, \alpha_s \,  \widetilde{\Phi}^1_{D/2}(\tilde{w})  \right) \\ & + a_2 \, \left( 2  \Phi^1_{D/2-1}(\tilde{w}) - \eta \, \alpha_s \,  \widetilde{\Phi}^1_{D/2-1}(\tilde{w}) \right)  \tfrac{\Kb^2}{k^2}  \Big] \, . 
\end{split}
\ee
The case with one potential insertion gives
\be\label{master1}
\begin{split}
{\rm Tr}\left[\, \cP^{-1} \,  \cV_1 \, \cP^{-1} \,  \p_t \cR_k \right] = & \,
\tfrac{k^{D}}{(4\pi)^{D/2}} \int_x \, a_0 \left( 2  \Phi^2_{D/2}(\tilde{w}) - \eta \, \alpha_s \,  \widetilde{\Phi}^2_{D/2}(\tilde{w})  \right)\tfrac{\Kb^2}{k^2} \, , \\
{\rm Tr}\left[\, \cP^{-1} \,  \cV_2 \, \cP^{-1} \,  \p_t \cR_k \right] = & \, - \, 
\tfrac{k^{D}}{(4\pi)^{D/2}} \int_x \, a_0 \left( 2  \Phi^2_{D/2}(\tilde{w}) - \eta \, \alpha_s \,  \widetilde{\Phi}^2_{D/2}(\tilde{w})  \right) \tfrac{\Kb^2}{k^2} \, , \\
{\rm Tr}\left[\, \cP^{-1} \, \cV_3 \, \cP^{-1} \,  \p_t \cR_k \right] = & \, 0 \, .
\end{split}
\ee
At the level of two insertions only the trace containing $(\cV_3)^2$ contributes to the flow. In this case, the application of off-diagonal heat-kernel techniques yields
\be\label{master2}
{\rm Tr}\left[(\cV_3)^2 \, \cP^{-3} \,  \p_t \cR_k \right] = - \tfrac{1}{2}
\tfrac{k^{D}}{(4\pi)^{D/2}} \int_x \, a_0 \left( 2  \Phi^2_{D/2+1}(\tilde{w}) - \eta \, \alpha_s \,  \widetilde{\Phi}^2_{D/2+1}(\tilde{w})  \right) \tfrac{\Kb^2}{k^2} \, .  
\ee
Here $a_0$ and $a_2$ are the spin-dependent heat-kernel coefficients introduced in App.\ \ref{App.A} and $\tilde{w} \equiv w k^{-2}$. Note that once a trace contains two derivatives of the background curvature, all remaining derivatives may be computed freely, since commutators give rise to terms which do not contribute to the flow of $G_k$ and $\Lambda_k$.
\subsection{Trace contributions in the gravitational sector}
At this stage, we have all the ingredients for evaluating the operator 
traces appearing on the right-hand-side of the FRGE, keeping all terms contributing to the truncation \eqref{defdimless}. In order to cast the resulting expressions into compact form, it is convenient to combine the threshold functions \eqref{thresholdfcts} according to
\be\label{qfct}
q_n^p(w) \equiv 2 \, \Phi^p_n(w) - \eta \, \widetilde{\Phi}^p_n(w) \, , 
\ee
and recall the definition of the dimensionless quantities \eqref{defdimless}. Moreover, all traces include the proper factors of $1/2$ and signs appearing on the right-hand-side of the FRGE.

We first evaluate the traces arising from the blocks of $\Gamma^{(2)} + \cR_k$ which are one-dimensional in field space. In the gravitational sector, this comprises the contributions of the component fields $h_{ij}, u_i, v_i$, and $B$. Applying the master formulas \eqref{master0} and \eqref{master1} and adding the results, one has
\be\label{tracegrav}
\begin{split}
	{\rm Tr}|_{hh} =  & \tfrac{k^D}{2 \, (4\pi)^{D/2}} 
	\int_x \Big[ 
	\tfrac{(d+1)(d-2)}{2} \, q^1_{D/2}(-2 \lambda) 
	 + \tfrac{d^4-2d^3-d^2+14d+36}{12d^2} \,  q^1_{D/2-1}(-2 \lambda) \, \tfrac{\Kb^2}{k^2} \\ & \qquad \qquad \quad
	- {\tfrac{(d-2)^2(d+1)^2}{2 \, d^2} } \, q^2_{D/2}(-2\lambda)  \tfrac{\Kb^2}{k^2} \
	\Big] \, , \\
	{\rm Tr}|_{vv} = & \tfrac{k^D}{2 \, (4\pi)^{D/2}} 
	\int_x \Big[ (d-1) \, q^1_{D/2}(-2 \lambda)  
	+  \tfrac{d^3-2d^2+d+6}{6d^2} \, q^1_{D/2-1}(-2 \lambda)\, \tfrac{\Kb^2}{k^2} \\ & \qquad \qquad  \quad
	- \tfrac{(d-1)(d^2-5d+7)}{d^2} \, q^2_{D/2}(-2\lambda) \, \tfrac{\Kb^2}{k^2} \
	\Big] \\
		{\rm Tr}|_{uu} = & \tfrac{k^D}{2 \, (4\pi)^{D/2}} 
		\int_x \Big[ (d-1) \, q^1_{D/2}(0)  
		+  \tfrac{d^3-2d^2+d+6}{6d^2} \, q^1_{D/2-1}(0)\, \tfrac{\Kb^2}{k^2} \\ & \qquad \qquad  \quad
		- \tfrac{(d-1)(d-2)}{d} \, q^2_{D/2}(0) \, \tfrac{\Kb^2}{k^2} \
		\Big] \\
 {\rm Tr}|_{BB} = & \tfrac{k^D}{2 \, (4\pi)^{D/2}} 
 \int_x \Big[ q^1_{D/2}(0)  
 +  \tfrac{d-1}{6d} \, q^1_{D/2-1}(0)\, \tfrac{\Kb^2}{k^2} 
 - \tfrac{(d-1)^2}{d^2} \, q^2_{D/2}(0) \, \tfrac{\Kb^2}{k^2} \
 \Big]
\end{split}
\ee

The evaluation of the traces in the ghost sector follows along the same lines. In this case one also has a contribution from the third master trace \eqref{master2}. The total contributions of the scalar ghosts is then given by
\be\label{gh1}
- {\rm Tr}|_{\bar{c}c} = - \tfrac{k^D}{(4\pi)^{D/2}} \int_x
\Big\{ 2 \, \Phi^1_{D/2} + \tfrac{\Kb^2}{k^2} \left[ \tfrac{d-1}{3d}  \Phi^1_{D/2-1} + 2  \Phi^1_{D/2} - \tfrac{4}{d^2}  \Phi^1_{D/2+1} \right] \Big\} \, , 
\ee
where all threshold functions are evaluated at zero argument.
Recalling that the vector ghost $b_i$ is not subject to a transverse constraint, the trace evaluates to
\be\label{gh2}
- {\rm Tr}|_{\bar{b}b} = - \tfrac{k^D}{(4\pi)^{D/2}} \int_x
\Big\{ 2d \, \Phi^1_{D/2} + \tfrac{\Kb^2}{k^2} \left[ \tfrac{d-1}{3} \, \Phi^1_{D/2-1} + \tfrac{8}{d} \, \Phi^1_{D/2} - \tfrac{4}{d} \, \Phi^1_{D/2+1}\right] \Big\}.
\ee

The last contribution of the flow is provided by the three scalar fields $\xi = (\Nh, E, \psi)$. Inspecting \eqref{eqS2Frank}, one finds that the block $\Gamma^{(2)} + \cR_k$ appearing in this sector is given by a $3\times 3$-matrix in field space with non-zero off-diagonal entries. Applying
the decomposition \eqref{PVdec} the matrix $\cP$ resulting from
\eqref{eqS2Frank} is
\be
\renewcommand{\arraystretch}{1.2}
\begin{split}
\cP = & (32 \pi G_k)^{-1}
\left[\begin{array}{ccc}
	\Delta_0            & \tfrac{1}{2} \left(\Delta_0 - 2 \Lambda\right)   & \tfrac{d-1}{2}  \left(\Delta_0 - 2 \Lambda\right)   \\
	\tfrac{1}{2} \left(\Delta_0 - 2 \Lambda\right) & \tfrac{1}{4} \left(\Delta_0 - 2 \Lambda\right)	   & - \tfrac{d-1}{4} \left(\Delta_0 - 2 \Lambda\right)      \\
	\; \tfrac{d-1}{2} \left(\Delta_0 - 2 \Lambda\right) \; & \; - \tfrac{d-1}{4} \left(\Delta_0 - 2 \Lambda\right) \; & \; - \tfrac{(d-1)(d-3)}{4} \left(\Delta_0 - 2 \Lambda\right) \;
\end{array}\right] \, , \\
\end{split}
\ee
while the matrix $\cV$ is symmetric with entries
\be
\begin{array}{ll}
	\cV_{11} =  - \tfrac{2(d-1)}{d^2} \left(2 \Kb^2 + d \dot{\Kb} \right) \, , \qquad & 
	\cV_{12} = - \tfrac{5d^2-12d+16}{8d^2} \Kb^2 	\\
\cV_{22} =  - \tfrac{d-1}{4d} \left(\Kb^2 + 2 \dot{\Kb} \right) \, , \quad &
\cV_{13} = - \tfrac{(d-1)(5d^2-12d+16)}{8d^2} \Kb^2
 \\
\cV_{33} =  \tfrac{(d-3)(d-1)^2}{4d} \left(\Kb^2 + 2 \dot{\Kb} \right) \, , \quad &
\cV_{23} = \tfrac{(d-1)^2}{4d} \, \left(\Kb^2 + 2 \dot{\Kb} \right) \, . 
\end{array}
\ee
Applying \eqref{Rscheme}, the cutoff $\cR_{k}$ in this sector is given by
\be
\renewcommand{\arraystretch}{1.2}
\cR_k = (32 \pi G_k)^{-1} \, R_k \, 
\left[\begin{array}{ccc}
1            & \tfrac{1}{2}    & \tfrac{d-1}{2}     \\
\tfrac{1}{2} & \tfrac{1}{4}	   & - \tfrac{d-1}{4}      \\
\; \tfrac{d-1}{2} \; & \; - \tfrac{d-1}{4} \; & \; - \tfrac{(d-1)(d-3)}{4} \;
\end{array}\right] \, . 
\ee
The master traces \eqref{master0} and \eqref{master1} also hold in the case where $\cP$ and $\cV$ are matrix valued. Constructing the inverse of $\cP$ on field space explicitly and evaluating the corresponding traces, the contribution of this block to the flow is found as
\be\label{scalars0}
\begin{split}
	{\rm Tr}|_{\xi\xi} = \tfrac{k^{D}}{2(4\pi)^{D/2}} & \int_x  \Big[ 
	2 \, q^1_{D/2}(-2\lambda) +  q^1_{D/2}\left(-\tfrac{d}{d-1}\lambda\right)   \\ & 
	+ \tfrac{d-1}{6d} \, \left( 2 \, q^1_{D/2-1}({-2\lambda}) + q^1_{D/2-1}\left(-\tfrac{d}{d-1}\lambda\right) \right)  \tfrac{\Kb^2}{k^2}   
	  \\
	 & - \left( \tfrac{2(d-1)}{d} \,   q^2_{D/2}\left(-2\lambda\right)  - \tfrac{3 d^3 + 6d^2-16d + 16}{4 d^2 (d-1)}
	 q^2_{D/2}\left(-\tfrac{d}{d-1}\lambda\right) \right) \tfrac{\Kb^2}{k^2}  
	 \Big] \, . 
\end{split}
\ee
The traces \eqref{tracegrav}, \eqref{gh1}, \eqref{gh2}, and \eqref{scalars0} complete the evaluation of the flow equation on a flat FRW background. Substituting these expressions into the FRGE \eqref{FRGE} and retaining the terms present in \eqref{FRGElhs} then leads to the beta functions \eqref{betafunction} where the threshold functions are evaluated with a Litim type regulator \eqref{thresholdfcts}.

\subsection{Minimally coupled matter fields}
\label{App.C3}
At the level of the Einstein-Hilbert truncation \eqref{GammaEH}, including the contribution of the matter sector \eqref{matter} to the flow of Newton's constant and the cosmological constant is rather straightforward. When expanding the matter fields around a vanishing background value, the Hessian $\Gamma^{(2)}$ arising in the matter sector contains variations with respect to the matter fields only and all Laplacians reduce to the background Laplacians. The resulting contributions of the matter trace are then identical to the ones obtained in the metric formulation \cite{Percacci:2002ie,Percacci:2003jz,Dona:2012am}. The trace capturing the contributions of the the $N_S$ scalar fields $\phi$ yields
\be\label{scalarmattertrace}
{\rm Tr}|_{\phi\phi} = N_S \, \tfrac{k^D}{(4\pi)^{D/2}} \int_x \Big\{ \Phi^1_{D/2}(0)  + \tfrac{d-1}{6d} \, \Phi^1_{D/2-1}(0) \, \tfrac{\Kb^2}{k^2} \Big\} \, . 
\ee
The gauge sector, comprising $N_V$ gauge fields $A_\mu$ and the corresponding Faddeev-Popov ghosts $\bar{C}, C$ contributes 
\be\label{Atrace}
{\rm Tr}|_{AA} = N_V \, \tfrac{k^D}{(4\pi)^{D/2}} \int_x \Big\{ (d+1) \, \Phi^1_{D/2}(0) + \tfrac{(d-1)(d^2+2d-11)}{6d \, (d+1)} \, \Phi^1_{D/2-1}(0) \,  \tfrac{\Kb^2}{k^2} \Big\} \, , 
\ee
and 
\be\label{CCbtrace}
- {\rm Tr}|_{\bar{C}C} = N_V \, \tfrac{k^D}{(4\pi)^{D/2}} \int_x \Big\{ 2 \, \Phi^1_{D/2}(0)  + \tfrac{d-1}{3d} \, \Phi^1_{D/2-1}(0) \,  \tfrac{\Kb^2}{k^2} \Big\} \, .
\ee
Adding eqs.\ \eqref{Atrace} and \eqref{CCbtrace} gives the total contribution of the gauge fields to the flow
\be\label{gaugemattertrace}
{\rm Tr}|_{\rm GF} = N_V \, \tfrac{k^D}{(4\pi)^{D/2}} \int_x \Big\{ (d-1) \, \Phi^1_{D/2}(0)  + \tfrac{(d-1)(d^2-13)}{6d \, (d+1)} \, \Phi^1_{D/2-1}(0) \,  \tfrac{\Kb^2}{k^2} \Big\} \, . 
\ee

When evaluating the contribution of the fermionic degrees of freedom, we follow the discussion \cite{Dona:2012am}, resulting in
\be\label{diracmattertrace}
{\rm Tr}|_{\psi\psi} = -  \tfrac{N_D \, 2^{(d+1)/2} \, k^D}{(4\pi)^{D/2}} \int_x \Big\{\Phi^1_{D/2}(0) +  \tfrac{d-1}{d} \left[ \left( \tfrac{1}{6} - \tfrac{r}{4}  \right) \Phi^1_{D/2-1}(0) - \tfrac{1-r}{4} \Phi^2_{D/2}(0) \right]  \tfrac{\Kb^2}{k^2} \Big\} \, . 
\ee
Here $r$ is a numerical coefficient which depends on the precise implementation of the regulating function: $r=0$ for a Type I regulator while the Type II construction of \cite{Dona:2012am} corresponds to $r=1$. In order to be consistent with the evaluation of the other traces in the gravitational and matter sectors, we will resort to the Type I regulator scheme, setting $r=1$.
Adding the results \eqref{scalarmattertrace}, \eqref{gaugemattertrace}, and \eqref{diracmattertrace} to the contribution from the gravitational sector gives rise to the $N_S$, $N_V$, and $N_D$-dependent terms in the beta functions \eqref{betafunction}.

\end{appendix}


\end{document}